\documentclass{aa}

\pdfoutput=1

\usepackage{amsmath}
\usepackage{bm}
\usepackage{color}
\usepackage{filecontents}
\usepackage{flushend}
\usepackage{graphicx}
\usepackage{hyperref}
\usepackage[all]{hypcap}
\usepackage{longtable}
\usepackage{multirow}
\usepackage{natbib}
\usepackage{pdfpages}
\usepackage{tablefootnote}
\usepackage{txfonts}
\usepackage[capitalize]{cleveref}

\def\fun#1#2{\lower3.6pt\vbox{\baselineskip0pt\lineskip.9pt
\ialign{$\mathsurround=0pt#1\hfil##\hfil$\crcr#2\crcr\sim\crcr}}}

\newcommand{\rv}{\mathbf{r}}
\newcommand{\kv}{\mathbf{k}}

\newcommand{\densshear}{\tilde{\gamma}^{I}}

\newcommand{\ebar}{\bar{\gamma}}

\usepackage{array}
\newcolumntype{L}[1]{>{\raggedright\let\newline\\\arraybackslash\hspace{0pt}}m{#1}}
\newcolumntype{C}[1]{>{\centering\let\newline\\\arraybackslash\hspace{0pt}}m{#1}}
\newcolumntype{R}[1]{>{\raggedleft\let\newline\\\arraybackslash\hspace{0pt}}m{#1}}

\usepackage{natbib}
\bibpunct{(}{)}{;}{a}{}{,} 

\hypersetup{colorlinks,
  urlcolor=blue,
  citecolor=blue}

\def\lcdm{$\Lambda$CDM}
\def\mass{$M_{200}$}
\def\meneacs{MENeaCS}
\def\mc{MENeaCS+CCCP}
\def\radius{$r_{200}$}
\def\sdelta{$\mathcal{S}_\Delta$}
\def\vdisp{$\sigma_{200}$}
\def\Ncl{90}
\def\Nmembers{14,576}
\def\Nwithinr{9,054}
\def\Nspec{38,104}
\def\tma{\tablefootmark{a}}
\def\tmb{\tablefootmark{b}}
\def\tmc{\tablefootmark{c}}

\def\exa{Experimental Astronomy}

\begin{document}

\title{Constraints on the alignment of galaxies in galaxy clusters from $\sim$14,000 spectroscopic 
members}

\author{Crist\'obal~Sif\'on\inst{1} \and
        Henk~Hoekstra\inst{1} \and
        Marcello~Cacciato\inst{1} \and
        Massimo~Viola\inst{1} \and
        Fabian~K\"ohlinger\inst{1} \and
        Remco~F.~J.~van~der~Burg\inst{1} \and
        David~J.~Sand\inst{2} \and
        Melissa~L.~Graham\inst{3}}

\institute{Leiden Observatory, Leiden University, PO Box 9513, NL-2300 RA Leiden, Netherlands
           \and
           Department of Physics, Texas Tech University, 2500 Broadway St., Lubbock, TX 79409, USA
           \and
           Astronomy Department, B-20 Hearst Field Annex \#3411,
           University of California at Berkeley, Berkeley, CA 94720-3411, USA
           }
\abstract{
Torques acting on galaxies lead to physical alignments, but the resulting ellipticity correlations 
are difficult to  predict. As they constitute a major contaminant for cosmic shear studies, it is 
important to constrain the intrinsic alignment signal observationally. We measured the alignments 
of satellite galaxies within \Ncl\  massive galaxy clusters in the redshift range $0.05<z<0.55$ and 
quantified their impact on the cosmic shear signal. We combined a sample of 38,104 galaxies with 
spectroscopic redshifts with high-quality data from the Canada-France-Hawaii Telescope. We used 
phase-space information to select \Nmembers\ cluster members, 14,250 of which have shape 
measurements and measured three different types of alignment: the radial alignment of satellite 
galaxies toward the brightest cluster galaxies (BCGs), the common orientations of satellite 
galaxies and BCGs, and the radial alignments of satellites with each other. Residual systematic 
effects are much smaller than the statistical uncertainties. We detect no galaxy alignment of any 
kind out to at least $3r_{200}$. The signal is consistent with zero for both blue and red galaxies, 
bright and faint ones, and also for subsamples of clusters based on redshift, dynamical mass, and 
dynamical state. These conclusions are unchanged if we expand the sample with bright cluster 
members from the red sequence. We augment our constraints with those from the literature to 
estimate the importance of the intrinsic alignments of satellites compared to those of central 
galaxies, for which the alignments are described by the linear alignment model. Comparison of the 
alignment signals to the expected uncertainties of current surveys such as the Kilo-Degree Survey 
suggests that the linear alignment model is an adequate treatment of intrinsic alignments, but it 
is not clear whether this will be the case for larger surveys.
}


\titlerunning{Intrinsic Alignments of Galaxies in Clusters}
\authorrunning{C.~Sif\'on et al.}

\maketitle

\section{Introduction}

Tidal torques tend to align triaxial satellite galaxies toward the center of the larger ``host'' 
gravitational potential as they orbit around its center. This mechanism is well established in 
numerical simulations, where galaxies are typically locked pointing toward the centers of 
clusters, possibly with brief periodic misalignments depending on the specific orbit, well within 
a Hubble time \citep[e.g.,][]{ciotti94,altay06,faltenbacher08,pereira08,pereira10}. In a 
hierarchical clustering scenario, this effect could be coupled with alignments arising from the 
nonlinear evolution of structure. Therefore the patterns and evolution of galaxy alignments---if 
any---contain important information about the initial conditions that gave rise to the present-day 
cosmic web, as well as the formation history and environments of galaxies.

Additionally, these galaxy alignments (commonly referred to as ``intrinsic,'' as opposed to 
apparent, alignments) are a potential contaminant of cosmic shear, which is a measurment of 
the coherent distortions of galaxies in the background of a matter distribution. While the signal 
from these intrinsic alignments is weak enough that it is not relevant for weak lensing 
measurements of galaxy clusters (and in general cluster members can be identified and removed to a 
sufficient level), it is a concern for large-area cosmic shear surveys, which are more susceptible 
to this contamination, and where the requirements on precision and accuracy are more stringent. The 
contamination induced by these galaxy alignments into cosmic shear measurements can be divided into 
two effects. The first effect is the tidal alignment of galaxies with similar formation histories, 
so-called intrinsic-intrinsic or II signal. Since this effect is restricted to pairs with common 
formation or evolutionary histories, this II signal can be avoided by selecting pairs of galaxies 
with large angular and/or redshift separations \citep[e.g.,][]{king02,heymans03,heymans04}. The 
second effect is more subtle and more difficult to control: the same gravitational field that 
aligns galaxies within a halo is responsible for the deflection of the light coming from background 
galaxies \citep{hirata04}. This effect is referred to as gravitational-intrinsic or GI signal (for 
consistency, the lensing signal itself is referred to as gravitational-gravitational, or GG, 
signal). In tomographic analyses, it is possible to account for this effect through its distinct 
redshift dependence \citep{king05,joachimi08,zhang10a,zhang10b} or, inversely, to measure it from 
cosmic shear data by boosting its signal while suppressing the contribution from gravitational 
lensing \citep{joachimi10b}. This cross-correlation has recently been shown to exist also between 
galaxy-galaxy lensing and CMB lensing \citep{hall14,troxel14}. Intrinsic alignments can also be 
modeled directly in cosmic shear data and marginalized over to extract cosmological parameters 
\citep{joachimi10a,heymans13}. In an attempt to identify a consistent model for galaxy shapes and 
alignments, \cite{joachimi13a,joachimi13b} have tried to match semi-analytical models to galaxies 
from the COSMOS survey \citep{scoville07} and find that the intrinsic alignment contamination to 
upcoming cosmic shear surveys should be $<10\%$.

Recent large photometric and spectroscopic surveys such as the 2-degree Field redshift survey 
\citep[2dF,][]{colless01} and the Sloan Digital Sky Survey \citep[SDSS,][]{york00} have allowed the 
study of galaxy alignments out to several tens of Mpc exploiting cross-correlation techniques, with 
robust direct detections of the GI signal up to $z\sim0.7$ between galaxy samples with large 
line-of-sight separations \citep{mandelbaum06,hirata07,joachimi11}, although \cite{mandelbaum11} 
reported a null detection. However, the II signal is much weaker than the GI signal in 
nontomographic studies at intermediate redshifts typical of these surveys, and has typically 
eluded detection \citep[e.g.,][]{mandelbaum06,mandelbaum11,blazek12}.

On smaller scales the history of these measurements goes back further, but the issue is far from 
settled. Early measurements of galaxy alignments focused on galaxy clusters, trying to understand
galaxy formation and (co-)evolution. \cite{rood72} were the first to claim a detection of a 
preferential direction of galaxies in clusters. Specifically, they found that satellite galaxies in 
Abell~2199 tend to point in the direction of the major axis of the BCG. However, most subsequent 
measurements have been consistent with random orientations of satellite galaxies in clusters 
\citep[e.g.,][]{hawley75,thompson76,dekel85,vankampen90,trevese92,panko09,hung12}, although some
authors have also claimed significant nonrandom orientations of these cluster satellites 
\citep[e.g.,][]{djorgovski83,godlowski98,godlowski10,baier03,plionis03}.

More recent studies have focused on smaller mass galaxy groups, where the number of objects is much 
larger. Similar to the results summarized above, most of these measurements are consistent with no 
alignments \citep{bernstein02,hao11,schneider13,chisari14}, although there are claims of 
significant detections \citep[e.g.,][]{pereira05,faltenbacher07}.\footnote{We discuss sources of 
this discrepancy in light of recent results in \cref{s:agreement}.} Interestingly, although this 
effect might be expected to be stronger for more massive halos, \cite{agustsson06} found that 
satellite galaxies are radially aligned in galaxy-scale halos. However, \cite{hao11} and 
\cite{schneider13} have shown that the results can depend on the method used to estimate the 
direction of the satellite galaxies, so each result must be taken with care.

In general, there is clear tension between observations and numerical N-body simulations, with 
the latter predicting much higher signals than have been observed. This discrepancy can be 
attributed, for instance, to a misalignment between stars and dark matter, such that stars---being 
more centrally concentrated than dark matter---react more slowly and less strongly to tidal 
torquing from the parent halo \citep{pereira10,tenneti14}. Whatever the physical reasons of this 
discrepancy, the potential impact of the choice of intrinsic alignment {\em model} on cosmological 
parameter estimation \citep{kirk12} makes it imperative that we know the level of intrinsic 
alignments to high precision at all relevant mass and spatial scales, and this can only be achieved 
through detailed observations.

In this work, we study the alignments of galaxies in clusters from a sample of galaxy clusters with 
high-quality photometric observations and a large number of spectroscopic redshifts from archival 
sources. We measure different kinds of alignments, assess systematic errors, and use the halo model 
to characterize galaxy alignments in the context of upcoming cosmic shear analyses.

We adopt a flat \lcdm\ cosmology with $\Omega_\Lambda=0.7$, $\Omega_M=0.3$ and
$H_0=70\,\mathrm{km\,s^{-1}\,Mpc^{-1}}$. To calculate the power spectra, we use $\sigma_8=0.8$, 
$\Omega_bh^2=0.0245$, and $n_s=1.0$. All magnitudes are \texttt{MAG\_AUTO} from SExtractor in the 
AB system, and all absolute magnitudes and luminosities are in the rest frame of the corresponding 
cluster. 

\section{Data}

\subsection{Cluster Sample and Photometry}\label{s:photometry}

\begin{figure}
 \centerline{\includegraphics[width=3.5in]{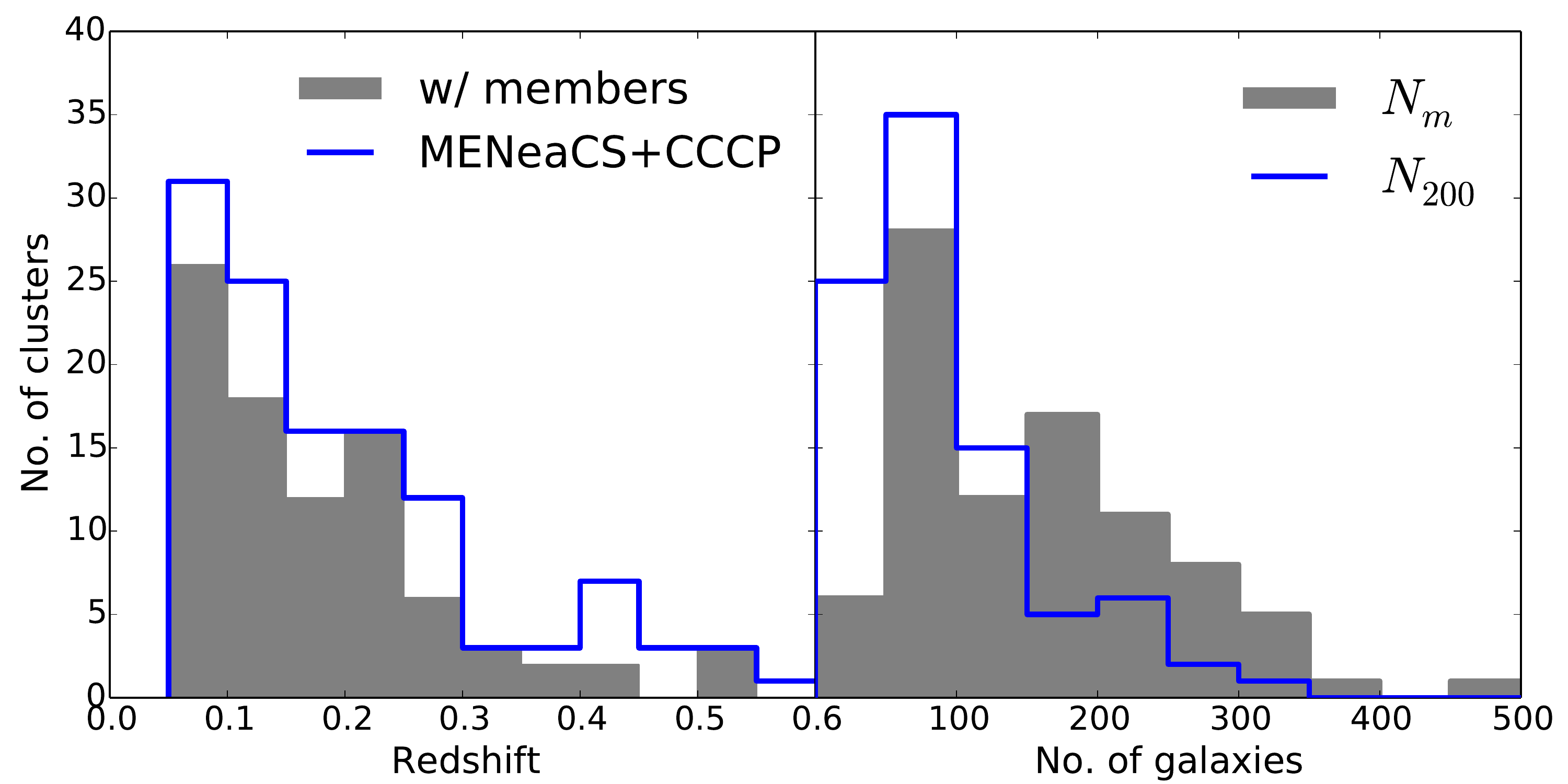}}
 \caption{\small \textit{Left:} redshift distributions of all \mc\ clusters (blue histogram) and 
clusters used in this study (gray filled histogram). \textit{Right:} distributions of number of 
spectroscopic members, $N_m$ (gray filled histogram), and number of spectroscopic members within 
\radius, $N_{200}$ (blue histogram). Abell~2142, with $N_m=1052$ and $N_{200}=731$, is not shown.}
\label{f:Nm}
\end{figure}

The cluster sample is drawn from two large, nonoverlapping X-ray selected cluster surveys carried 
out with the Canada-France-Hawaii Telescope (CFHT), namely the Multi-Epoch Nearby Cluster Survey 
\citep[MENeaCS;][]{sand12} and the Canadian Cluster Comparison Project \citep[CCCP;][]{hoekstra12}. 
MENeaCS performed multi-epoch observations of 57 clusters in the redshift range $0.05<z<0.15$, 
aimed at measuring the supernova Ia rate in these clusters. For this, clusters were observed using 
the $g$ and $r$ bands with MegaCam. CCCP was designed to study the scaling relations between 
different tracers of mass in galaxy clusters, and includes 50 clusters in the redshift range 
$0.15<z<0.55$. Of these, 20 clusters had archival {\it B}- and {\it R}-band data taken with the 
CFH12k camera, and 30 clusters were observed with the $g$ and $r$ bands with MegaCam 
\citep{hoekstra07,hoekstra12}.

The data were reduced using the Elixir pipeline \citep{magnier04}, and processed further following 
the method outlined in \cite{vanderburg13}, which is summarized below. In order to simplify point 
spread function (PSF) modeling for shape measurements, we homogenized the PSF across the entire 
field-of-view by finding a locally-varying convolution kernel that makes the PSF circular and 
Gaussian everywhere \citep{kuijken08}. This PSF homogenization was done for each exposure, after 
which the individual exposures were co-added. By applying a Gaussian weight function to measure 
aperture fluxes we optimized color measurements in terms of S/N (see \citealt{kuijken08} and 
\citealt{vanderburg13}, Appendix A). This gaussianization process introduces correlations in pixel 
noise, which we do not account for. As we show in \cref{s:systematics}, this is not a problem for 
the present analysis.

We performed object detection with SExtractor \citep{bertin96} in dual-image mode, using the $r$ 
(or $R$) band as detection image. All sources were detected in the $r$-band image obtained by 
stacking the nonhomogenized images. Photometric zero points were calibrated using the stellar 
locus regression (SLR) software developed by 
\cite{kelly14}\footnote{\url{https://code.google.com/p/big-macs-calibrate/}}. SLR uses the known 
colors of stars to obtain solutions for the photometric zero points of any photometric catalog, 
correcting for instrumental response and atmospheric and Galactic extinctions \citep[see 
also][]{high09}. We used the Two-Micron All Sky Survey \citep[2MASS,][]{skrutskie06} $J$-band star 
catalog in addition to our MegaCam $g$ and $r$, or CFH12k $B$ and $R$, observations as inputs to 
the SLR. We retrieved extinction values in the $J$ band from the NASA/IPAC Extragalactic Database 
(NED)\footnote{\url{http://ned.ipac.caltech.edu/}}, which use the reddening measurements of 
\cite{schlafly11}. With these two colors, plus the absolute photometric calibrations of 2MASS 
(including the extinction correction), we obtained absolute zero points for the CFHT catalogs. For 
clusters within the SDSS footprint we also use the SDSS \textit{griz} photometry to check for 
consistency, from which we conclude that SLR-corrected zero points are calibrated to an absolute 
uncertainty of $\sim\!0.01$ mag. Galaxies were separated from stars by visual inspection of the 
magnitude-size\footnote{Here, sizes are given by \texttt{FLUX\_RADIUS} from SExtractor.} plane for 
each cluster individually. Stars occupy a well-defined region in this plane, having essentially a 
single size up to the saturation flux. Given that the stacks generally have a sub-arcsecond sized 
PSF, galaxies used here are large compared to the PSF and are therefore easily distinguishable from 
stars.

We computed $r$-band absolute magnitudes, $M_r$, using 
EzGal\footnote{\url{http://www.baryons.org/ezgal/}} \citep{mancone12}, using a passive evolution 
Charlot \& Bruzual 2007 \citep[unpublished, see][]{bruzual03} single burst model with solar 
metallicity and formation redshift $z_f=5$.

\subsection{Spectroscopic Data}

\begin{figure}
 \centerline{\includegraphics[width=3.5in]{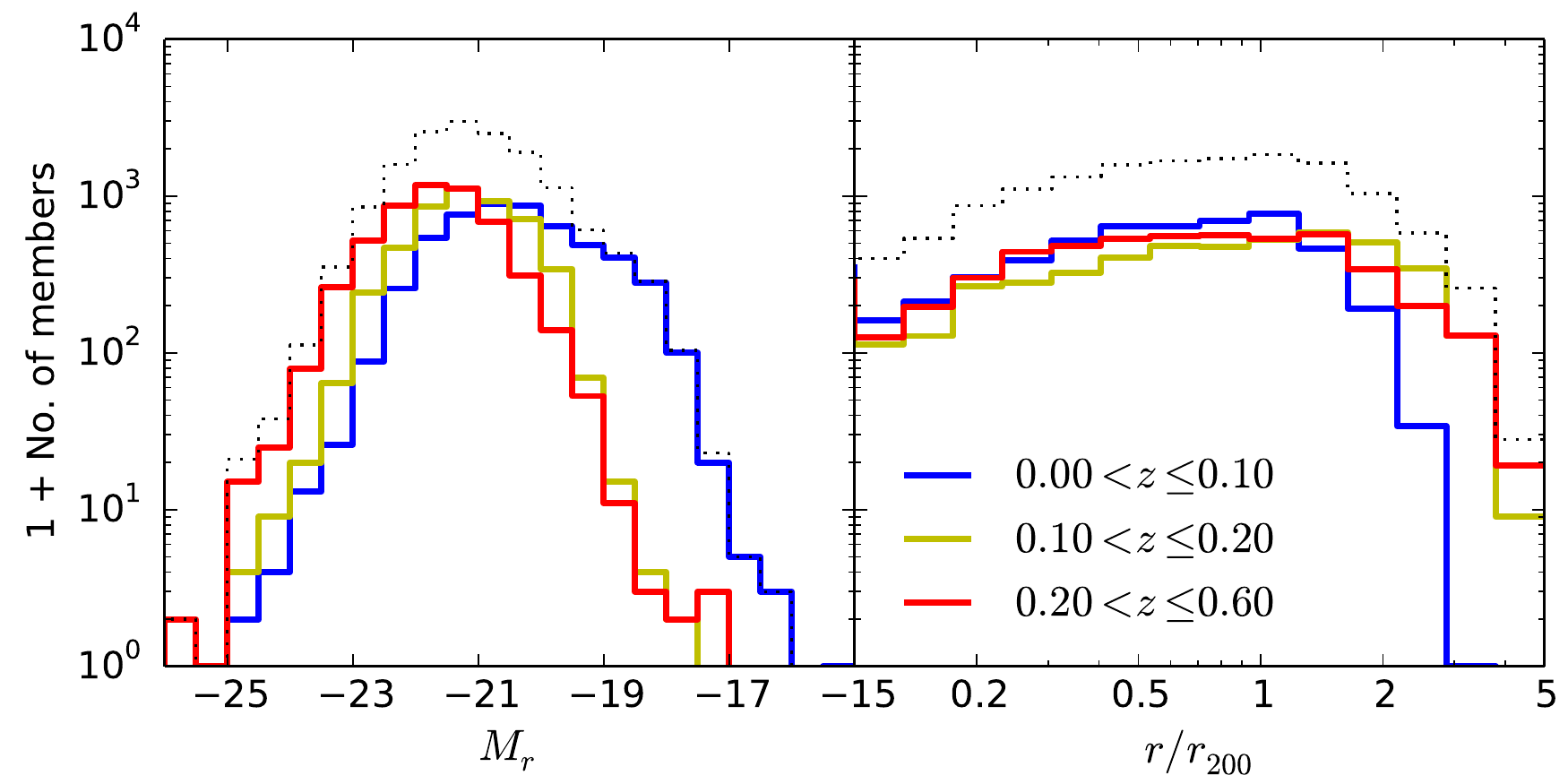}}
\caption{\small Distribution of spectroscopic members. \textit{Left:} as a function of rest-frame 
absolute {\it r}-band magnitude. \textit{Right:} as a function of cluster-centric distance in units 
of \radius. The dotted black lines show the distribution of the full sample; the solid lines show 
the distribution split into three redshift bins of approximately equal number of clusters.}
\label{f:coverage}
\end{figure}

\begin{table}
\begin{center}
\caption{Spectroscopic catalogs.}
\label{t:catalogs}
\begin{tabular}{l r@{}l r@{}l c c}
\hline\hline
Source & \multicolumn{2}{c}{Total} & \multicolumn{2}{c}{Unique}  & Clusters & Avg.\ unique \\
       & \multicolumn{2}{c}{redshifts} & \multicolumn{2}{c}{redshifts} &  & per cluster \\[0.5ex]
\hline
\multicolumn{7}{c}{Compilations} \\
NED          &   16, & 125    &  9, & 161 &      79 & 116 \\
WLTV         &    1, & 613    &  1, & 399 & \,\,\,2 & 700 \\
CNOC         &    1, & 427    &  1, & 266 &      10 & 127 \\
SDSS DR10    &   14, & 634    & 13, & 995 &      62 & 226 \\
HeCS         &    8, & 470    &  8, & 368 &      27 & 310 \\
MENeaCS-spec &    1, & 966    &  1, & 966 &      12 & 164 \\
\hline
\multicolumn{7}{c}{Single-cluster} \\
\multicolumn{3}{l}{\cite{geller14}} &  & 834 & \,\,\,1 & 834 \\
\multicolumn{3}{l}{\cite{ebeling14}} & 1, & 115 & \,\,\,1 & 1,115\,\,\,\, \\[0.2ex]
\hline
{\bf Total}  &  & & {\bf 38,} & {\bf 104} & {\bf \Ncl} & {\bf 423} \\
\hline
\end{tabular}
\end{center}
\end{table}

We searched for spectroscopic redshifts around all clusters in the \mc\ sample in six archival 
sources: NED, the WIYN Long-Term Variability survey \citep[WLTV;][]{crawford11}, the Canadian 
Network for Observational Cosmology Survey \citep[CNOC;][]{yee96,yee98,ellingson97,abraham98}, the 
SDSS Data Release 10\footnote{\url{http://www.sdss3.org/dr10/data_access/}} \citep[DR10;][]{ahn13} 
which is part of SDSS-III \citep{sdss3} and the Hectospec Cluster Survey \citep[HeCS;][]{rines13}. 
We also include redshifts from the \meneacs\ spectroscopic survey (hereafter MENeaCS-spec). When a 
galaxy was found in more than one of these catalogs, each catalog replaced the preceding as listed 
above and in \Cref{t:catalogs}. Thus we included all redshifts from MENeaCS-spec. In NED and in 
SDSS DR10, we searchws for galaxies with spectroscopic redshifts within a radius of 1 deg.\ around 
each cluster, but only galaxies in our photometric catalogs were included in the analysis. From 
NED, we discarded all flagged redshifts (i.e., all those whose \texttt{Redshift Flag} field is not 
blank) and kept only redshifts with at least 4 significant digits to ensure that only spectroscopic 
redshifts with enough precision are included. From SDSS we only included galaxies with 
\texttt{zWarning=0}. The NED search includes redshifts obtained as part of large surveys such as 
the 2dF, the 2MASS spectroscopic survey (2MRS), the WIde-field Nearby Galaxy cluster Survey 
(WINGS), and the WiggleZ Dark Energy Survey.

Additionally, we included the redshift catalogs of Abell~383 and MACS~J0717.5+3745 recently 
published by \cite{geller14} and \cite{ebeling14}, respectively, which total 1,949 redshifts within 
our CFHT images. From the catalog of \cite{ebeling14} we used only redshifts with quality flag 1 or 
2. We also highlight the redshift catalog of Abell~2142 by \cite{owers11}, containing $\sim\!1,800$ 
galaxies, which is included as part of the NED catalog.

\Cref{t:catalogs} lists the number of (unique) spectroscopic redshifts included from each catalog 
and from how many cluster fields they are taken. The largest redshift catalog(s) for each cluster 
are listed in \Cref{t:dynamics}, including the largest catalogs within NED; for most clusters the 
NED redshifts come mainly from one or two large catalogs (with $\gtrsim90\%$ of redshifts). The 
final spectroscopic sample is summarized in the last row of \Cref{t:catalogs}: it contains \Nspec\ 
redshifts in the direction of \Ncl\ clusters, selected to have at least 30 members, at least 10 of 
which must be within \radius\ (see \cref{s:specmembers}). The left panel of \cref{f:Nm} shows the 
redshift distribution of these \Ncl\ clusters, compared to the entire \mc\ sample. The analysis in 
this paper refers only to these \Ncl\ clusters, which are listed in \Cref{t:dynamics}.

\section{Galaxy Samples}

\begin{table*}
\begin{center}
\caption{Cluster sample, redshifts, and velocity dispersions.}
\label{t:dynamics}
\begin{tabular}{L{3.1cm} c r r r@{$\,\pm\,$}l r@{$\,\pm\,$}l r@{$\,\pm\,$}l r@{}l c c}
\hline\hline
Cluster      & $z$ & $N_m$ & $N_{200}$ & \multicolumn{2}{c}{\vdisp} & 
\multicolumn{2}{c}{\mass} & \multicolumn{2}{c}{\radius} & \multicolumn{2}{c}{\sdelta} & Main & Main 
NED \\
 &  &  &  & \multicolumn{2}{c}{$(\mathrm{km\,s^{-1}})$} & \multicolumn{2}{c}{$(10^{14}\,M_\odot)$} 
& 
\multicolumn{2}{c}{$(\mathrm{Mpc})$} &  &  & sources\tma & sources\tmb \\[0.5ex]
\hline
Abell~85       & 0.0555 &  284 &  248 &  967 &  55 & 10.0 & 1.7 & 2.03 & 0.12 & 
$0.01_{-0.00}^{+0.00}$ & (D) &  3 & 14 \\[0.4ex]
Abell~115      & 0.1930 &   73 &   73 & 1028 & 108 & 11.2 & 3.5 & 2.03 & 0.21 & 
$0.24_{-0.07}^{+0.02}$ & (R) &  -- & 1 \\[0.4ex]
Abell~119      & 0.0443 &  268 &  255 &  875 &  48 &  7.5 & 1.2 & 1.85 & 0.10 &  $<0.01$ & (D) & 3 
& 7 \\[0.4ex]
Abell~133      & 0.0558 &   62 &   59 &  791 &  79 &  5.5 & 1.7 & 1.66 & 0.17 & 
$0.24_{-0.14}^{+0.03}$ & (R) &  -- & 21 \\[0.4ex]
Abell~209\tmc  & 0.2090 &  110 &  110 & 1170 &  99 & 16.4 & 4.1 & 2.28 & 0.19 & 
$0.01_{-0.01}^{+0.00}$ & (D) &  -- & 27 \\[0.4ex]
Abell~222\tmc  & 0.2132 &   76 &   76 &  881 &  79 &  7.0 & 1.9 & 1.72 & 0.15 & 
$0.30_{-0.04}^{+0.04}$ & (R) &  -- & 11,33 \\[0.4ex]
Abell~223\tmc  & 0.2076 &   64 &   64 &  910 &  80 &  7.8 & 2.1 & 1.78 & 0.16 & 
$0.05_{-0.02}^{+0.02}$ & (D) &  -- & 11,33 \\[0.4ex]
Abell~267      & 0.2291 &  219 &  156 & 1006 &  74 & 10.3 & 2.3 & 1.95 & 0.14 &  $<0.01$ & (D) & 
3,4 & -- \\[0.4ex]
Abell~2670     & 0.0763 &  241 &  196 &  919 &  46 &  8.5 & 1.3 & 1.91 & 0.10 &  $<0.01$ & (D) & 3 
& 21 \\[0.4ex]
Abell~2703     & 0.1140 &   75 &   13 &  657 &  53 &  3.1 & 0.8 & 1.35 & 0.11 & 
$0.17_{-0.05}^{+0.18}$ & (R) &  3 & -- \\[0.4ex]
Abell~383      & 0.1885 &  182 &  134 &  918 &  53 &  8.1 & 1.4 & 1.81 & 0.11 &  $<0.01$ & (D) & -- 
& 18 \\[0.4ex]
Abell~399      & 0.0718 &  250 &  229 & 1046 &  47 & 12.5 & 1.7 & 2.18 & 0.10 &  $<0.01$ & (D) & 5 
& 20 \\[0.4ex]
Abell~401      & 0.0735 &  104 &   83 &  933 &  81 &  8.9 & 2.3 & 1.94 & 0.17 & 
$0.32_{-0.05}^{+0.19}$ & (R) &  -- & 20 \\[0.4ex]
Abell~520      & 0.2007 &  153 &  127 & 1045 &  73 & 11.8 & 2.5 & 2.05 & 0.14 & 
$0.27_{-0.11}^{+0.04}$ & (R) &  -- & 19 \\[0.4ex]
Abell~521\tmc  & 0.2469 &   95 &   95 & 1002 &  95 & 10.1 & 2.9 & 1.92 & 0.18 & 
$0.06_{-0.02}^{+0.02}$ & (D) &  -- & 16 \\[0.4ex]
Abell~545\tmc  & 0.1577 &   80 &   80 & 1038 &  89 & 11.8 & 3.0 & 2.08 & 0.18 & 
$0.07_{-0.00}^{+0.02}$ & (R) &  -- & 2 \\[0.4ex]
Abell~553      & 0.0670 &   54 &   44 &  665 &  75 &  3.3 & 1.1 & 1.39 & 0.16 & 
$0.01_{-0.00}^{+0.03}$ & (D) &  5 & -- \\[0.4ex]
Abell~586      & 0.1704 &   33 &   21 &  803 & 104 &  5.4 & 2.1 & 1.60 & 0.21 & 
$0.28_{-0.07}^{+0.12}$ & (R) &  3 & -- \\[0.4ex]
Abell~644\tmc  & 0.0696 &   31 &   31 &  625 &  96 &  2.7 & 1.2 & 1.31 & 0.20 & 
$0.67_{-0.05}^{+0.11}$ & (R) &  -- & 24 \\[0.4ex]
Abell~646      & 0.1266 &  259 &   69 &  707 &  66 &  3.8 & 1.1 & 1.44 & 0.13 & 
$0.07_{-0.04}^{+0.22}$ & (D) &  3,4 & -- \\[0.4ex]
Abell~655      & 0.1271 &  306 &  109 &  938 &  57 &  8.8 & 1.6 & 1.91 & 0.12 & 
$0.03_{-0.02}^{+0.01}$ & (D) &  3,4 & -- \\[0.4ex]
Abell~697      & 0.2821 &  215 &  106 & 1161 &  89 & 15.4 & 3.6 & 2.19 & 0.17 & 
$0.44_{-0.02}^{+0.05}$ & (R) &  3,4 & -- \\[0.4ex]
Abell~754      & 0.0546 &  305 &  300 &  959 &  43 &  9.8 & 1.3 & 2.01 & 0.09 & 
$0.01_{-0.00}^{+0.00}$ & (D) &  -- & 8 \\[0.4ex]
Abell~780\tmc  & 0.0547 &   33 &   33 &  822 & 113 &  6.2 & 2.5 & 1.73 & 0.24 & 
$0.01_{-0.00}^{+0.01}$ & (D) &  -- & 21 \\[0.4ex]
Abell~795      & 0.1385 &  166 &  117 &  768 &  59 &  4.9 & 1.1 & 1.56 & 0.12 & 
$0.06_{-0.02}^{+0.07}$ & (D) &  3,4,5 & -- \\[0.4ex]
Abell~851      & 0.4038 &   53 &   47 &  999 & 138 &  9.3 & 3.8 & 1.77 & 0.24 & 
$0.01_{-0.01}^{+0.03}$ & (D) &  -- & 3,12 \\[0.4ex]
Abell~959      & 0.2882 &   67 &   67 &  982 & 101 &  9.4 & 2.9 & 1.85 & 0.19 & 
$0.08_{-0.03}^{+0.04}$ & (D) &  -- & 5 \\[0.4ex]
Abell~961      & 0.1275 &   58 &   16 &  740 & 142 &  4.4 & 2.5 & 1.51 & 0.29 & 
$0.84_{-0.13}^{+0.06}$ & (R) &  3 & -- \\[0.4ex]
Abell~963      & 0.2036 &  165 &   85 &  922 &  64 &  8.1 & 1.7 & 1.81 & 0.13 & 
$0.17_{-0.06}^{+0.06}$ & (R) &  3,4 & -- \\[0.4ex]
Abell~990      & 0.1421 &  209 &   86 &  829 &  96 &  6.1 & 2.1 & 1.68 & 0.19 & 
$0.74_{-0.05}^{+0.03}$ & (R) &  3,4,5 & -- \\[0.4ex]
Abell~1033     & 0.1224 &  170 &   98 &  762 &  52 &  4.8 & 1.0 & 1.56 & 0.11 & 
$0.59_{-0.21}^{+0.08}$ & (R) &  3,4 & -- \\[0.4ex]
Abell~1068     & 0.1393 &  104 &   32 &  740 & 160 &  4.3 & 2.8 & 1.50 & 0.32 & 
$0.93_{-0.04}^{+0.04}$ & (R) &  3,4 & -- \\[0.4ex]
Abell~1132     & 0.1349 &  160 &   55 &  727 &  89 &  4.1 & 1.5 & 1.48 & 0.18 & 
$0.13_{-0.05}^{+0.06}$ & (R) &  3,4 & -- \\[0.4ex]
Abell~1234     & 0.1638 &   54 &   30 &  513 &  86 &  1.4 & 0.7 & 1.03 & 0.17 & 
$0.87_{-0.07}^{+0.06}$ & (R) &  3,4 & -- \\[0.4ex]
Abell~1246     & 0.1920 &  207 &   87 &  956 &  84 &  9.1 & 2.4 & 1.89 & 0.17 & 
$0.01_{-0.00}^{+0.01}$ & (D) &  3,4 & -- \\[0.4ex]
Abell~1285     & 0.1078 &   77 &   51 &  826 &  90 &  6.1 & 2.0 & 1.70 & 0.19 & 
$0.52_{-0.18}^{+0.08}$ & (R) &  5 & -- \\[0.4ex]
Abell~1361     & 0.1157 &  143 &   46 &  587 &  62 &  2.2 & 0.7 & 1.21 & 0.13 & 
$0.39_{-0.07}^{+0.15}$ & (R) &  3,4 & -- \\[0.4ex]
Abell~1413     & 0.1418 &  124 &   65 &  881 &  81 &  7.3 & 2.0 & 1.78 & 0.16 & 
$0.72_{-0.23}^{+0.07}$ & (R) &  3,4 & -- \\[0.4ex]
Abell~1650     & 0.0841 &  266 &  140 &  720 &  48 &  4.1 & 0.8 & 1.50 & 0.10 & 
$0.99_{-0.02}^{+0.00}$ & (R) &  3,5 & -- \\[0.4ex]
Abell~1651     & 0.0847 &  214 &  138 &  903 &  51 &  8.0 & 1.4 & 1.87 & 0.11 & 
$0.65_{-0.16}^{+0.04}$ & (R) &  -- & 9,21 \\[0.4ex]
Abell~1689     & 0.1847 &  252 &  235 & 1429 &  81 & 30.0 & 5.1 & 2.82 & 0.16 &  $<0.01$ & (D) & 
3,4 & -- \\[0.4ex]
Abell~1758     & 0.2772 &  133 &   34 &  744 & 107 &  4.1 & 1.8 & 1.41 & 0.20 & 
$0.11_{-0.03}^{+0.06}$ & (R) &  3,4 & -- \\[0.4ex]
Abell~1763     & 0.2323 &  186 &  103 & 1130 &  81 & 14.6 & 3.1 & 2.18 & 0.16 &  $<0.01$ & (D) & 
3,4 & -- \\[0.4ex]
Abell~1781     & 0.0622 &   54 &   16 &  419 &  93 &  0.8 & 0.5 & 0.88 & 0.19 & 
$0.92_{-0.19}^{+0.02}$ & (R) &  3 & -- \\[0.4ex]
Abell~1795     & 0.0629 &  191 &  133 &  778 &  51 &  5.2 & 1.0 & 1.63 & 0.11 & 
$0.26_{-0.09}^{+0.04}$ & (R) &  3 & 21 \\[0.4ex]
Abell~1835     & 0.2506 &  195 &   41 &  762 & 106 &  4.5 & 1.9 & 1.46 & 0.20 & 
$0.66_{-0.23}^{+0.06}$ & (R) &  3,4 & -- \\[0.4ex]
Abell~1914     & 0.1671 &  257 &  146 &  911 &  54 &  7.9 & 1.4 & 1.82 & 0.11 & 
$0.86_{-0.05}^{+0.03}$ & (R) &  3,4 & -- \\[0.4ex]
Abell~1927     & 0.0953 &  138 &   58 &  725 &  58 &  4.2 & 1.0 & 1.50 & 0.12 & 
$0.25_{-0.08}^{+0.02}$ & (R) &  3,5 & -- \\[0.4ex]
\hline
\end{tabular}
\end{center}
\end{table*}

\addtocounter{table}{-1}
\begin{table*}
\begin{center}
\caption{{\it Continued}}
\begin{tabular}{L{3.1cm} c r r r@{$\,\pm\,$}l r@{$\,\pm\,$}l r@{$\,\pm\,$}l r@{}l c c}
\hline\hline
Cluster      & $z$ & $N_m$ & $N_{200}$ & \multicolumn{2}{c}{\vdisp} & 
\multicolumn{2}{c}{\mass} & \multicolumn{2}{c}{\radius} & \multicolumn{2}{c}{\sdelta} & Main & Main 
NED \\
 &  &  &  & \multicolumn{2}{c}{$(\mathrm{km\,s^{-1}})$} & \multicolumn{2}{c}{$(10^{14}\,M_\odot)$} 
& \multicolumn{2}{c}{$(\mathrm{Mpc})$} &  &  & sources\tma & sources\tmb \\ 
\hline
Abell~1942     & 0.2257 &   51 &   27 &  820 & 140 &  5.6 & 2.9 & 1.59 & 0.27 & 
$0.04_{-0.01}^{+0.05}$ & (D) &  3 & -- \\[0.4ex]
Abell~1991     & 0.0587 &  175 &   99 &  553 &  45 &  1.9 & 0.5 & 1.17 & 0.10 & 
$0.12_{-0.05}^{+0.09}$ & (R) &  3,5 & -- \\[0.4ex]
Abell~2029     & 0.0777 &  317 &  181 & 1152 &  58 & 16.6 & 2.5 & 2.39 & 0.12 &  $<0.01$ & (D) & 3 
& 21 \\[0.4ex]
Abell~2033     & 0.0796 &  190 &   88 &  911 &  69 &  8.3 & 1.9 & 1.89 & 0.14 & 
$0.03_{-0.01}^{+0.03}$ & (D) &  3 & 21 \\[0.4ex]
Abell~2050     & 0.1202 &  164 &   82 &  854 &  80 &  6.7 & 1.9 & 1.74 & 0.16 & 
$0.34_{-0.03}^{+0.06}$ & (R) &  3,4 & -- \\[0.4ex]
Abell~2055     & 0.1028 &  154 &   69 &  697 &  64 &  3.7 & 1.0 & 1.44 & 0.13 & 
$0.04_{-0.00}^{+0.02}$ & (D) &  3,4 & 21 \\[0.4ex]
Abell~2064     & 0.0734 &   62 &   32 &  675 & 108 &  3.4 & 1.6 & 1.41 & 0.22 & 
$0.40_{-0.05}^{+0.13}$ & (R) &  3 & -- \\[0.4ex]
Abell~2065     & 0.0725 &  219 &  164 & 1095 &  67 & 14.3 & 2.6 & 2.28 & 0.14 & 
$0.03_{-0.01}^{+0.01}$ & (D) &  3 & -- \\[0.4ex]
Abell~2069     & 0.1139 &  331 &  146 &  966 &  63 &  9.7 & 1.9 & 1.98 & 0.13 & 
$0.01_{-0.00}^{+0.00}$ & (D) &  3,4 & -- \\[0.4ex]
Abell~2104     & 0.1547 &   90 &   56 & 1081 & 126 & 13.3 & 4.6 & 2.17 & 0.25 & 
$0.22_{-0.09}^{+0.09}$ & (R) &  -- & 23 \\[0.4ex]
Abell~2111     & 0.2281 &  256 &   83 &  738 &  66 &  4.1 & 1.1 & 1.43 & 0.13 & 
$0.46_{-0.04}^{+0.09}$ & (R) &  3,4 & 29 \\[0.4ex]
Abell~2125     & 0.2466 &  141 &   55 &  857 & 122 &  6.4 & 2.7 & 1.65 & 0.23 & 
$0.46_{-0.26}^{+0.01}$ & (R) &  -- & 28 \\[0.4ex]
Abell~2142     & 0.0903 & 1052 &  731 & 1086 &  31 & 13.9 & 1.2 & 2.24 & 0.06 & 
$0.01_{-0.00}^{+0.00}$ & (D) &  3 & 31 \\[0.4ex]
Abell~2163     & 0.2004 &  309 &  290 & 1279 &  53 & 21.5 & 2.7 & 2.51 & 0.10 & 
$0.03_{-0.01}^{+0.02}$ & (D) &  -- & 25 \\[0.4ex]
Abell~2204     & 0.1507 &  100 &   15 &  782 & 278 &  5.1 & 5.4 & 1.58 & 0.56 & 
$0.35_{-0.12}^{+0.33}$ & (R) &  -- & 32 \\[0.4ex]
Abell~2219     & 0.2255 &  364 &  241 & 1189 &  65 & 17.0 & 2.8 & 2.30 & 0.13 &  $<0.01$ & (D) & 
3,4 & 4 \\[0.4ex]
Abell~2259     & 0.1602 &  158 &   77 &  901 &  70 &  7.7 & 1.8 & 1.80 & 0.14 & 
$0.04_{-0.01}^{+0.12}$ & (D) &  3,4 & -- \\[0.4ex]
Abell~2261     & 0.2257 &  206 &   76 &  882 &  86 &  7.0 & 2.0 & 1.71 & 0.17 & 
$0.03_{-0.02}^{+0.05}$ & (D) &  3,4 & -- \\[0.4ex]
Abell~2319\tmc & 0.0538 &   83 &   83 & 1101 &  99 & 14.7 & 4.0 & 2.31 & 0.21 & 
$0.52_{-0.16}^{+0.07}$ & (R) &  -- & 30 \\[0.4ex]
Abell~2390     & 0.2287 &  136 &   92 & 1120 & 113 & 14.3 & 4.3 & 2.17 & 0.22 & 
$0.23_{-0.10}^{+0.01}$ & (R) &  2 & -- \\[0.4ex]
Abell~2409     & 0.1454 &  101 &   46 &  826 &  94 &  6.0 & 2.0 & 1.67 & 0.19 & 
$0.16_{-0.03}^{+0.09}$ & (R) &  3,5 & -- \\[0.4ex]
Abell~2440\tmc & 0.0906 &   88 &   88 &  766 &  61 &  4.9 & 1.2 & 1.59 & 0.13 & 
$0.13_{-0.02}^{+0.03}$ & (R) &  3 & 26 \\[0.4ex]
Abell~2495     & 0.0790 &   98 &   46 &  631 &  55 &  2.8 & 0.7 & 1.32 & 0.12 & 
$0.23_{-0.01}^{+0.06}$ & (R) &  3,5 & -- \\[0.4ex]
Abell~2537     & 0.2964 &  175 &   65 &  909 &  85 &  7.4 & 2.1 & 1.70 & 0.16 & 
$0.05_{-0.01}^{+0.04}$ & (D) &  3 & 6 \\[0.4ex]
Abell~2597     & 0.0829 &   39 &   17 &  682 & 131 &  3.5 & 2.0 & 1.42 & 0.27 & 
$0.37_{-0.09}^{+0.15}$ & (R) &  -- & 9,13 \\[0.4ex]
CL~0024.0+1652 & 0.3948 &  229 &  131 &  757 &  48 &  4.1 & 0.8 & 1.35 & 0.09 &  $<0.01$ & (D) & -- 
& 10 \\[0.4ex]
MACS~J0717.5+3745 & 0.5436 &  468 &  215 & 1370 &  79 & 22.0 & 3.8 & 2.24 & 0.13 &  $<0.01$ & (D) & 
-- & 15 \\[0.4ex]
MKW3S          & 0.0444 &  125 &   82 &  592 &  49 &  2.3 & 0.6 & 1.25 & 0.11 & 
$0.83_{-0.11}^{+0.06}$ & (R) &  3 & -- \\[0.4ex]
MS~0015.9+1609 & 0.5475 &  232 &  122 & 1330 & 115 & 20.0 & 5.2 & 2.17 & 0.19 & 
$0.32_{-0.18}^{+0.10}$ & (R) &  1,2 & -- \\[0.4ex]
MS~0440.5+0204 & 0.1962 &   51 &   35 &  742 & 103 &  4.3 & 1.8 & 1.46 & 0.20 & 
$0.26_{-0.13}^{+0.14}$ & (R) &  2 & -- \\[0.4ex]
MS~0451.6$-$0305 & 0.5382 &  247 &  200 & 1252 &  55 & 16.8 & 2.2 & 2.05 & 0.09 &  $<0.01$ & (D) & 
1,2 & -- \\[0.4ex]
MS~1008.1$-$1224 & 0.3077 &   86 &   85 & 1028 &  92 & 10.6 & 2.8 & 1.91 & 0.17 & 
$0.59_{-0.11}^{+0.07}$ & (R) &  2 & 22 \\[0.4ex]
MS~1224.7+2007 & 0.3258 &   33 &   29 &  790 &  95 &  4.8 & 1.7 & 1.46 & 0.17 & 
$0.57_{-0.07}^{+0.23}$ & (R) &  2,3 & -- \\[0.4ex]
MS~1231.3+1542 & 0.2347 &   84 &   65 &  710 &  57 &  3.7 & 0.9 & 1.38 & 0.11 & 
$0.87_{-0.03}^{+0.03}$ & (R) &  2,3 & -- \\[0.4ex]
MS~1358.4+6245 & 0.3289 &  189 &  152 & 1021 &  56 & 10.3 & 1.7 & 1.88 & 0.10 & 
$0.01_{-0.00}^{+0.01}$ & (D) &  2 & 17 \\[0.4ex]
MS~1455.0+2232 & 0.2565 &   57 &   57 &  928 & 111 &  8.0 & 2.9 & 1.77 & 0.21 & 
$0.58_{-0.24}^{+0.07}$ & (R) &  2,3 & -- \\[0.4ex]
MS~1512.4+3647 & 0.3719 &   30 &   29 &  960 & 170 &  8.4 & 4.4 & 1.73 & 0.30 & 
$0.04_{-0.00}^{+0.03}$ & (D) &  2 & -- \\[0.4ex]
MS~1621.5+2640 & 0.4254 &   70 &   41 &  724 &  82 &  3.5 & 1.2 & 1.27 & 0.14 & 
$0.17_{-0.04}^{+0.20}$ & (R) &  2 & -- \\[0.4ex]
\hline
\end{tabular}
\end{center}
\end{table*}

\addtocounter{table}{-1}
\begin{table*}
\begin{center}
\caption{{\it Continued}}
\begin{tabular}{L{3.1cm} c r r r@{$\,\pm\,$}l r@{$\,\pm\,$}l r@{$\,\pm\,$}l r@{}l c c}
\hline\hline
Cluster      & $z$ & $N_m$ & $N_{200}$ & \multicolumn{2}{c}{\vdisp} & 
\multicolumn{2}{c}{\mass} & \multicolumn{2}{c}{\radius} & \multicolumn{2}{c}{\sdelta} & Main & Main 
NED \\
 &  &  &  & \multicolumn{2}{c}{$(\mathrm{km\,s^{-1}})$} & \multicolumn{2}{c}{$(10^{14}\,M_\odot)$} 
& \multicolumn{2}{c}{$(\mathrm{Mpc})$} &  &  & sources\tma & sources\tmb \\ 
\hline
RX~J0736.6+3924 & 0.1179 &   62 &   28 &  432 &  64 &  0.9 & 0.4 & 0.89 & 0.13 & 
$0.52_{-0.03}^{+0.12}$ & (R) &  3,5 & -- \\[0.4ex]
ZwCl~0628.1+2502 & 0.0814 &   72 &   66 &  843 &  96 &  6.6 & 2.2 & 1.75 & 0.20 & 
$0.04_{-0.01}^{+0.02}$ & (D) &  5 & -- \\[0.4ex]
ZwCl~1023.3+1257 & 0.1425 &   84 &   24 &  622 & 108 &  2.6 & 1.3 & 1.26 & 0.22 & 
$0.79_{-0.17}^{+0.17}$ & (R) &  3,4 & -- \\[0.4ex]
ZwCl~1215.1+0400 & 0.0773 &  183 &  107 &  902 &  65 &  8.0 & 1.7 & 1.88 & 0.14 & 
$0.29_{-0.04}^{+0.02}$ & (R) &  3 & -- \\[0.4ex]
\hline
\end{tabular}
\end{center}
\tablefoot{
Columns are: (1) cluster name; (2): cluster redshift; (3): number of members out to 
arbitrary radius; (4): number of members within $r_{200}$; (5): velocity dispersion of members 
within $r_{200}$; (6): total mass within $r_{200}$; (7): cluster radius $r_{200}$; (8): 
significance level of the DS test, the letter in parenthesis shows whether a cluster is classified 
as disturbed (D), relaxed (R), or intermediate (I).\\
\tablefoottext{a}{Numbers are: (1): WLTV; (2): CNOC; (3): SDSS; (4): HeCS; (5): MENeaCS-spec. See 
text for references.}
\tablefoottext{b}{Catalogs extracted from NED that contribute significantly to each cluster.
These are: 
(1): \cite{barrena07}; (2): \cite{barrena11}; (3): \cite{belloni96}; (4): \cite{boschin04};
(5): \cite{boschin09}; (6): \cite{braglia09}; (7): WINGS \citep{cava09};
(8): \cite{christlein03}; (9): 2dF \citep{colless03}; (10): \cite{czoke01}; (11): \cite{dietrich02};
(12): \cite{dressler92}; (13): WiggleZ \citep{drinkwater10}; (14): \cite{durret98};
(15): \cite{ebeling14}; (16): \cite{ferrari03}; (17): \cite{fisher98}; (18): \cite{geller14};
(19): \cite{girardi08}; (20): \cite{hill93}; (21): 2MRS \citep{huchra12}; (22): \cite{jager04};
(23): \cite{liang00}; (24): \cite{martini07}; (25): \cite{maurogordato08};
(26): \cite{maurogordato11}; (27): \cite{mercurio03}; (28): \cite{miller04}; (29): \cite{miller06}; 
(30): \cite{oegerle95}; (31): \cite{owers11}; (32): \cite{pimbblet06}; (33): \cite{proust00}.}
\tablefoottext{c}{Spectroscopic members extend out to less than $0.8r_{200}$.}
}
\end{table*}

\subsection{Spectroscopic Members and Dynamical Masses}\label{s:specmembers}

\begin{figure}[h!]
 \centerline{\includegraphics[width=3.4in]{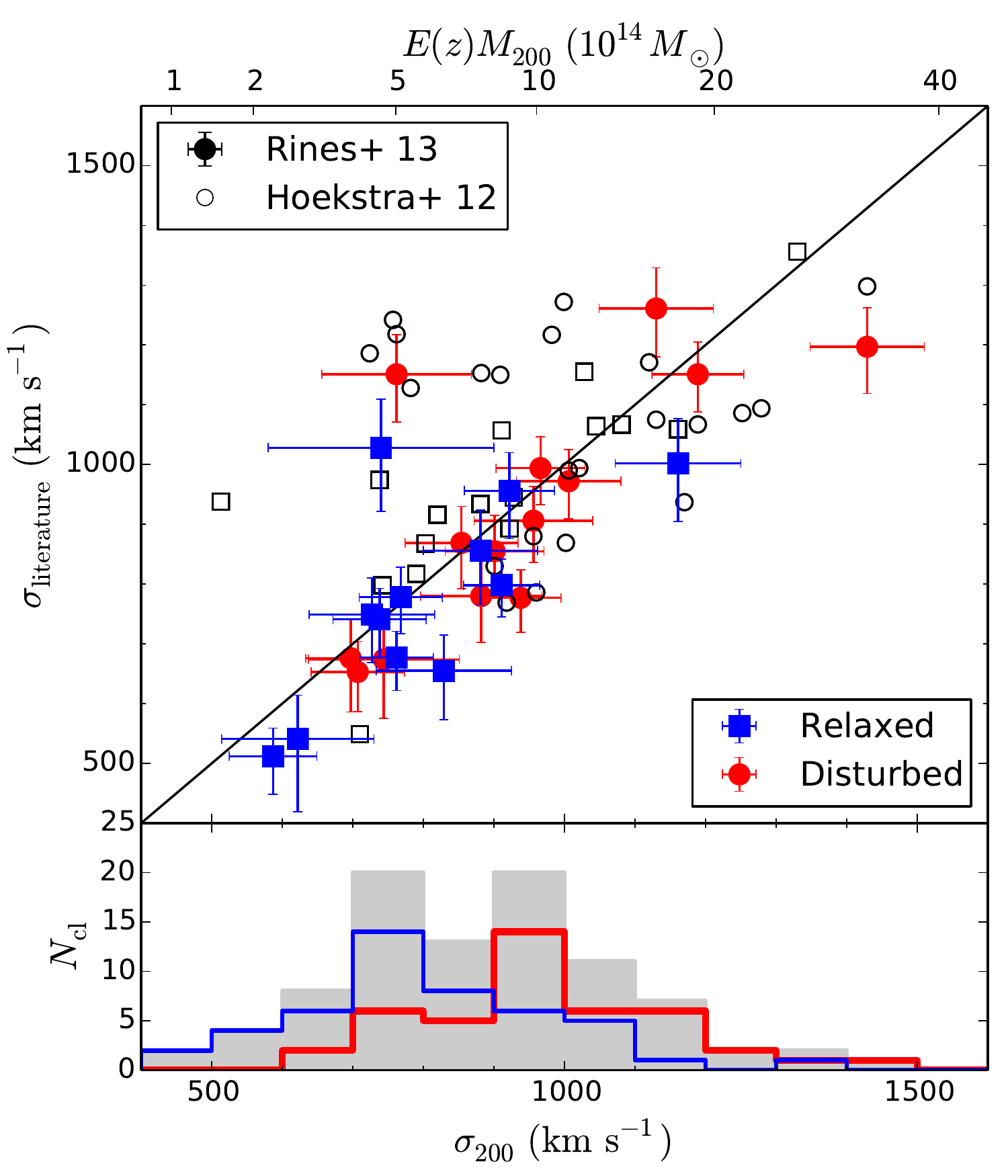}}
\caption{\small {\it Top:} comparison between velocity dispersions calculated from spectroscopic 
members in this work with those in \cite{rines13}, and with velocity dispersions calculated by 
fitting a single isothermal sphere to the weak lensing profile \citep{hoekstra12} for CCCP clusters 
used in this work. Errorbars are not shown for the latter for clarity. Squares and circles show 
relaxed and disturbed clusters, respectively. The black line shows the one-to-one relation, and the 
top axis shows $E(z)M_{200}$ for a given $\sigma_{200}$ from the \cite{evrard08} relation. {\it 
Bottom:} distribution of velocity dispersions of the full sample. The gray histogram shows the 
total distribution, with the blue and red histograms showing the distributions for relaxed and 
disturbed clusters, respectively.}
\label{f:comparesigma}
\end{figure}

Spectroscopic membership is determined using the shifting gapper method \citep{fadda96}, slightly 
adjusted from the implementation of \cite{sifon13}. We bin galaxies in radial bins of at least 15 
galaxies and 250 kpc and, for each radial bin, members are selected as those galaxies that are 
within $800\,\mathrm{km\,s^{-1}}$ from the main body of galaxies, which is bound by gaps of
$400\,\mathrm{km\,s^{-1}}$. The reduction in the velocity gaps compared to \cite{sifon13}---who 
used 1000 and 500 $\mathrm{km\,s^{-1}}$, respectively---is due to the larger number of galaxies 
used here, producing a distribution that has less obvious gaps in velocity space. In some cases, we 
introduced a radial cut determined from visual inspection. The left panel of \cref{f:coverage} 
shows that the distribution of confirmed cluster members is similar at low and high redshift for 
luminous ($M_r\lesssim-21$) galaxies but low-luminosity galaxies used here come mainly from low 
redshift clusters. We iteratively measure the velocity dispersion, \vdisp, as the biweight 
estimator of scale \citep{beers90} using all galaxies within \radius, correcting for measurement 
errors \citep{danese80}. Since the measurement uncertainties are not available for all galaxies 
(most notably, they are not given in NED), we use a fiducial value of $150\,\mathrm{km\,s^{-1}}$ 
for the uncertainty of all redshifts, which is a conservative estimate for recent measurements, but 
can be representative of older or low resolution measurements listed in NED. The change in mass 
introduced by this correction is, in any case, at the percent level for a cluster with 
$\sigma\sim1000\,\mathrm{km\,s^{-1}}$. The cluster redshift is determined iteratively in this 
process as the biweight estimator of location, considering again galaxies within \radius. We 
estimate the mass within \radius, namely \mass, using the $\sigma_{200}-M_{200}$ relation of 
\cite{evrard08}, and estimate $r_{200}=\left[3/(4\pi) \cdot M_{200}/(200\rho_c)\right]^{1/3}$. The 
resulting \vdisp\ is insensitive to uncertainties in $r_{200}$; varying \radius\ by $\pm20\%$ from 
those obtained from this relation only changes \vdisp\ by $\lesssim5\%$ for every cluster (in other 
words, velocity dispersion profiles are very close to flat near $r_{200}$). The uncertainties in 
the velocity dispersion are obtained from 1,000 jackknife samples drawn from all galaxies with 
peculiar velocities up to 3 times the cluster velocity dispersion; therefore quoted uncertainties 
include an estimate of the effect of membership selection. Uncertainties in the dynamical mass are 
obtained by propagating the uncertainties on the velocity dispersion. The dynamical properties 
described above are listed in \Cref{t:dynamics}, together with the number of members, $N_m$, and 
the number of members within \radius, $N_{200}$. The right panel of \cref{f:Nm} shows the 
distribution of $N_{200}$ and $N_m$, the number of members out to arbitrary radius.

In cases where the spectroscopic members do not reach out to \radius, we cannot infer \vdisp\ 
directly from the data. We therefore apply a correction to the measured velocity dispersion 
assuming the isotropic velocity dispersion profile of \cite{mamon10} and the mass--concentration 
relation of \cite{duffy08} to get the theoretical expectation for \vdisp\ given $\sigma(<r_{\rm 
max})$. This correction is linear with $r_{\rm max}$ for $r_{\rm max} \gtrsim 0.2 r_{200}$, with 
correction factors 0.93 and 0.81 for the velocity dispersion and mass, respectively, for $r_{\rm 
max}=0.6r_{200}$ (i.e., the mass within $0.6r_{200}$ is $0.81M_{200}$), only weakly dependent on 
mass and redshift. In our sample there are 14 clusters with $r_{\rm max}<r_{200}$, with a median of 
$r_{\rm max}/r_{200}=0.69$ and 10th and 90th percentiles of 0.51 and 0.79, respectively. For these 
clusters, we list the corrected values in \Cref{t:dynamics}.

There are a total of \Nmembers\ cluster members among \Ncl\ clusters, \Nwithinr\ of which are 
within \radius. The radial distribution of cluster members is shown in the right panel of 
\cref{f:coverage}. The spectroscopic members on average follow a Navarro-Frenk-White 
\citep[NFW,][]{navarro95} profile with concentration $c_{200}\sim 2$ (van der Burg et al., in 
prep). The median redshift of the sample is $z=0.144$, and the median velocity dispersion is 
$\sigma_{200}=881\,\mathrm{km\,s^{-1}}$ which translates to a median dynamical mass 
$M_{200}=7.2\times10^{14}\,M_\odot$ and a median size $r_{200}\sim1.7\,{\rm Mpc}$. The distribution 
of \vdisp\ is shown in the bottom panel of \cref{f:comparesigma}.

In \cref{f:comparesigma}, we compare the present velocity dispersions to those of \cite{rines13}; 
there is a large overlap between the two data sets (see \Cref{t:catalogs}). The two sets of 
measurements are consistent, with a median ratio $\langle\sigma_{200}/\sigma_{\rm 
Rines+13}\rangle=1.04\pm0.03$. We also show, for comparison, the singular isothermal sphere 
velocity dispersions fit by \cite{hoekstra12} to the weak lensing signal of 39 overlapping 
clusters, which are also consistent with our measurements to within 2\% on average. It is apparent 
from \cref{f:comparesigma} that the agreement between $\sigma_{200}$ and $\sigma_{\rm WL}$ is 
better 
for relaxed clusters than for disturbed ones, consistent with expectations. For comparison, using 
the velocity gaps used by \cite{sifon13} in our analysis, we obtain velocity dispersions larger 
than those of \cite{rines13} by $\sim$14\%.

\subsubsection{Dynamical State}\label{s:substructure}

\begin{figure*}
 \centerline{\includegraphics[width=2.4in]{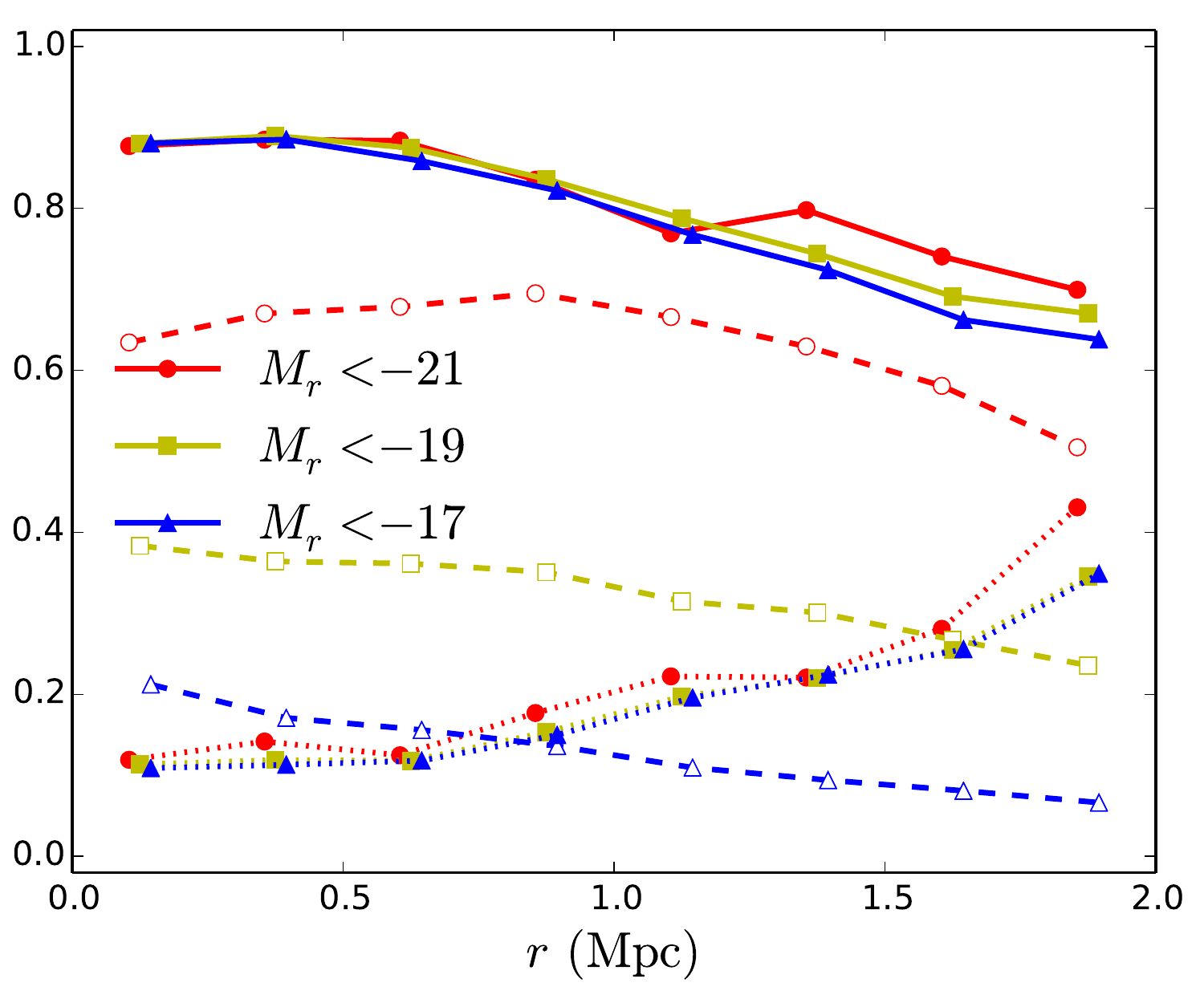}
             \includegraphics[width=2.4in]{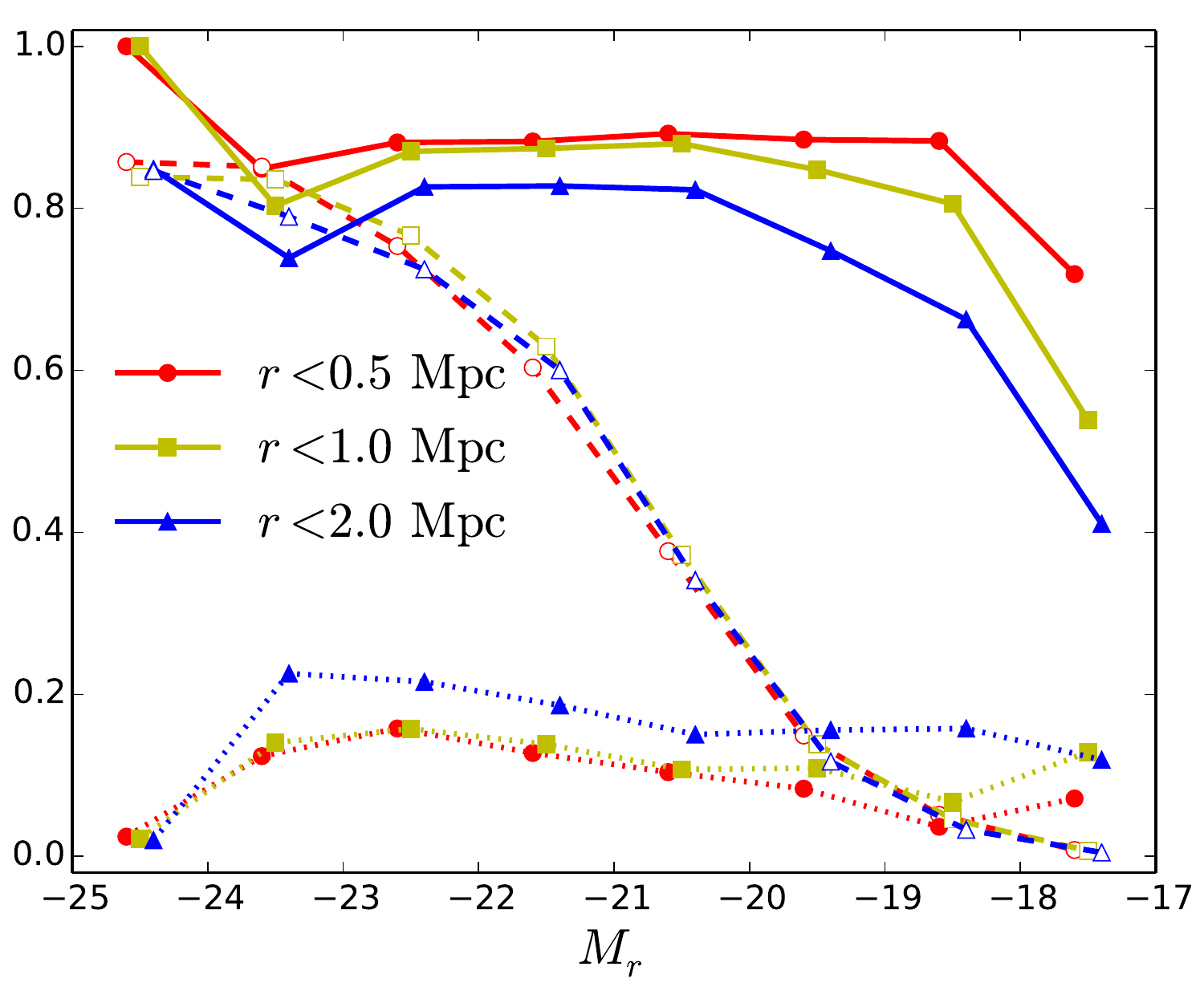}
             \includegraphics[width=2.4in]{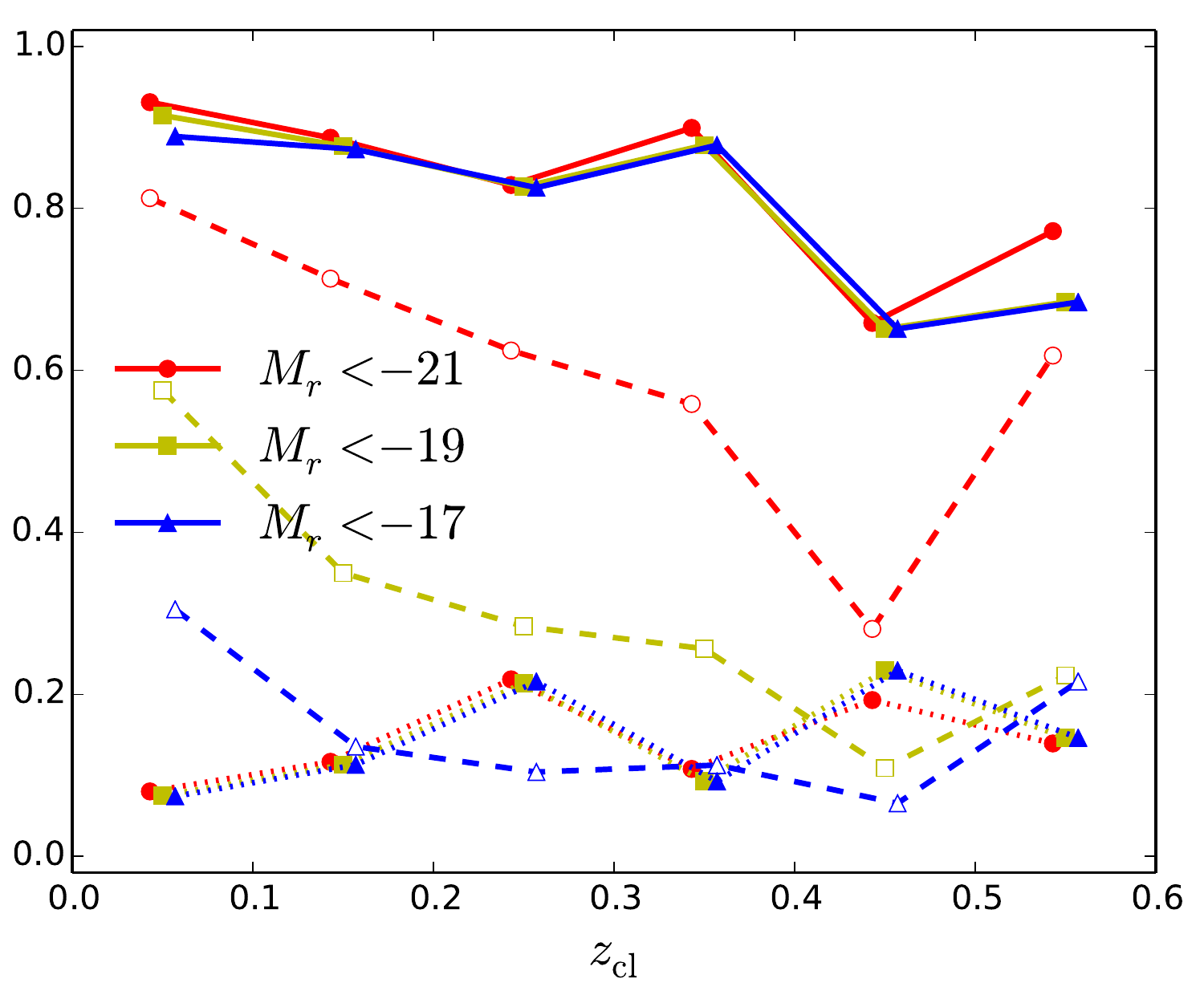}}
 \centerline{\includegraphics[width=6in]{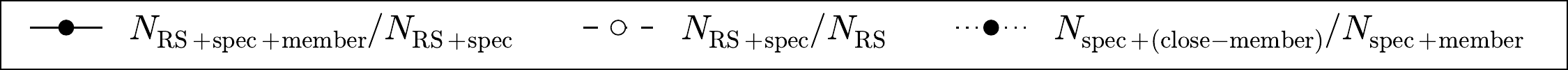}}
\caption{\small Purity of the red sequence (filled symbols with solid lines) and spectroscopic 
completeness within the red sequence (open symbols with dashed lines). Filled symbols with dotted 
lines show the fraction of galaxies that are not selected as members but that are within $\Delta 
z=0.03(1+z)$ of the cluster, which represents the contamination in an unbiased photometric redshift 
selection. \textit{Left:} as a function of cluster-centric distance, for different luminosity 
limits. \textit{Middle:} as a function of absolute magnitude, for different radial apertures. 
\textit{Right:} as a function of redshift for different luminosity limits, at an aperture of 1 Mpc. 
Note that points within a given line are independent (each line is a differential distribution), 
but lines of the same type are not independent from each other.}
\label{f:rspurity}
\end{figure*}

We can take further advantage of our large spectroscopic catalogs by studying the dynamical states 
of clusters. To this end we use the DS test \citep{dressler88}, which uses both the positions and 
velocities of galaxies. The DS test gives a measure of substructure by identifying galaxies that do 
not follow the cluster velocity distribution through the metric
\begin{equation}\label{eq:ds}
 \delta^2 = \frac{N_{\rm local}}{\sigma^2}\left[\left(\bar v_{\rm local} - \bar v\right)^2 + 
\left(\sigma_{\rm local} - \sigma\right)^2\right]^2\,,
\end{equation}
where $\bar v_{\rm local}$ and $\sigma_{\rm local}$ are the local velocity and velocity dispersion, 
measured for the $N_{\rm local}$ nearest neighbors around a test member, and $\bar v$ and $\sigma$ 
are the global values. The $\Delta$-statistic is the sum of $\delta$'s over all cluster members. 
This statistic is then measured 5,000 times after shuffling the velocities of cluster members, 
keeping positions fixed, with the same $N_{\rm local}$. The substructure significance (hereafter 
\sdelta) is the fraction of random samples which have $\Delta$ higher than that of the cluster. 
Errorbars are 68\% ranges obtained from 5 runs for each cluster, varying the number of neighbors 
within $\sqrt{N_{200}}-2 \leq N_{\rm local} \leq \sqrt{N_{200}}+2$, and the central value is their 
median. We run the DS test using only members within \radius\ because \radius\ is very close to the 
virial radius, beyond which the cluster \textit{should not} be relaxed, by definition.

The DS test is not designed to assess the dynamical state of clusters in general but specifically 
to find substructure, which furthermore has to have a different spatial and velocity location from 
the cluster itself. It is therefore incomplete; there are indeed examples of known merging clusters 
from which the DS test cannot find indications of substructure, most notably mergers along the 
plane of the sky \citep[e.g.,][]{menanteau12,barrena13}. This is the case here with Abell~520 
\citep[e.g.,][]{jee14}, for example. By means of $N$-body simulations, \cite{pinkney96} showed that 
$\mathcal{S}_\Delta<0.05$ is a reasonable condition to define a pure, but not necessarily complete, 
sample of dynamically disturbed clusters. We follow the results of \cite{pinkney96} in a 
conservative way, selecting as disturbed all cluster that are consistent with 
$\mathcal{S}_\Delta\leq0.05$ within errorbars (52 clusters). All others are classified as relaxed 
(38 clusters). This is conservative in the sense that we aim to construct a pure sample of relaxed 
clusters (see \cref{s:radial}). \Cref{t:dynamics} lists \sdelta\ together with the classification 
for each cluster. We find that more massive clusters tend to be classified as disturbed (D), while 
less massive clusters are generally relaxed (R); this can be seen in \cref{f:comparesigma}.

\subsection{Red Sequence Members}\label{s:photmembers}

While spectroscopy provides a clean sample of member galaxies from precise velocities, it suffers 
from incompleteness mainly due to two practical reasons: 1) obtaining a redshift for a galaxy is 
expensive; typically it takes $\sim\!30$ minutes of observations for galaxies in low-redshift 
($z\lesssim0.5$) clusters, depending on the telescope and observing conditions; and 2) only a 
limited number of galaxies can be targeted in a single observation because of slit overlap or fiber 
collisions.

Being a distinct feature of clusters, the red sequence provides an ideal complement to 
spectroscopic members. As we show below, for luminous galaxies near the centers of clusters this 
also provides a clean membership selection, though not as clean as spectroscopy. Using the red 
sequence {\it in addition to} the spectroscopic selection ensures that only a small fraction of 
galaxies need to be included through this more uncertain method, making the purity of the sample 
very close to 100\%.

To find the red sequence in each cluster, we first separate blue and red galaxies by fitting two 
one-dimensional gaussians to the color distribution of galaxies using an Error-Corrected Gaussian 
Mixture Model \citep[ECGMM,][]{hao09}. We then fit a straight line in color-magnitude space to the 
red galaxies using a maximum likelihood approach that accounts for intrinsic scatter and the 
measurement uncertainty in color, iteratively rejecting $2\sigma$ outliers. Details will be 
presented in a forthcoming paper (Sif\'on et al., in prep).

We assess the purity of the red sequence as a cluster member selection procedure by looking at red 
sequence members that have redshifts. There are in total 57,885 red sequence galaxies up to 
$M_r=-17$ and within 2 Mpc, 7,224 of which have a redshift measurement ($\sim$12\%). 
\cref{f:rspurity} shows that the red sequence is a high-fidelity member selection method even to 
large radii. Only the sample of both low-luminosity ($M_r\gtrsim-19$) and distant ($r\gtrsim1$ Mpc) 
red sequence members has a lower purity, although the latter is still $\gtrsim70\%$ for most of 
this distance-luminosity space. We include in the extended sample all red sequence galaxies more 
luminous than $M_r=-19$ within 1 Mpc of the cluster center. Within these parameter boundaries, 84\% 
of red sequence galaxies with a spectroscopic redshift are confirmed cluster members. This level of 
contamination (16\% of red sequence members) has no effects on our results.

The rightmost panel of \cref{f:rspurity} shows that up to $z\sim0.4$, the purity is extremely high 
($\sim$90\%) but then decreases to $\sim\!70\%$, because above $z\sim0.4$ the 4000\AA\ break 
is no longer bracketed by the $g$ and $r$ bands (similarly for the $B$ and $R$ bands). The 
completeness of the spectroscopic samples does drop noticeably with redshift because of the higher 
difficulty posed by spectroscopic observations of high redshift clusters.

\begin{figure}
 \centerline{\includegraphics[width=3.2in]{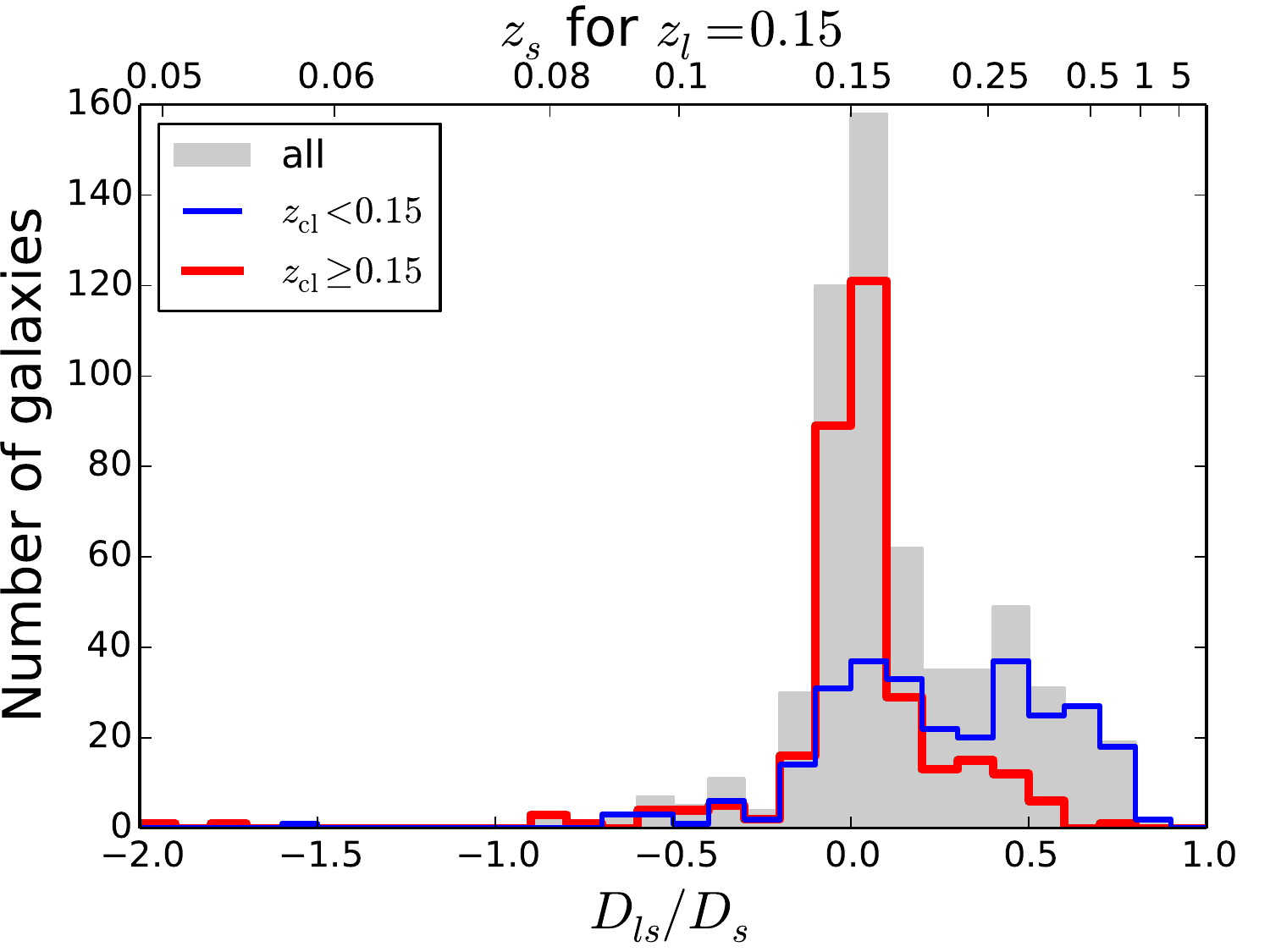}}
\caption{\small Distribution of the distance ratio, $D_{ls}/D_s$, for red sequence members that are 
confirmed to be nonmembers of the clusters from spectroscopic redshifts. The gray filled histogram 
shows red sequence galaxies from all clusters; the blue and red (empty) histograms show the 
distributions for clusters at low and high redshift, respectively. For illustration, the top axis 
shows the source redshift for a cluster at $z=0.15$.}
\label{f:rsbeta}
\end{figure}

From a lensing perspective, one other important ingredient in assessing the red sequence is the 
redshift distribution of the contaminating fraction, which we can quantify using red sequence 
galaxies that are confirmed to be outside the cluster. If they are sufficiently far behind the 
cluster, they could in fact be lensed, inducing a signal that we wish to avoid. If instead they are 
either very close behind or in front of the cluster, then they will not be lensed and will only 
dilute the signal. \cref{f:rsbeta} shows the redshift distribution of red sequence galaxies 
that are confirmed to be nonmembers through the distance ratio, $D_{ls}/D_s$, where $D_{ls}$ is 
the angular diameter distance between the lens (i.e., cluster) and the source galaxy, and $D_s$ is 
the angular diameter distance to the source. Note that $D_{ls}<0$ for $z_s<z_l$. Galaxies behind 
the clusters are lensed, resulting in (apparent) tangential alignments. The amplitude of this 
effect can be quantified by the ``lensing efficiency,'' $\beta$, defined as
\begin{equation}\label{eq:beta}
 \beta \equiv \left\langle\max\left(0, \frac{D_{ls}}{D_s}\right)\right\rangle\,,
\end{equation}
which is typically calculated using photometric redshifts or an average redshift distribution. Note 
that \cref{eq:beta} naturally accounts for galaxies in front of the cluster, which do not contain a 
lensing signal (but do introduce noise), which is especially important when using a generic 
redshift 
distribution, or full photometric redshift probability distributions (in which case background 
galaxies have a nonzero probability of being in front of the cluster). Of the 688 confirmed 
nonmembers in the red sequence, 496 (72\%) are behind the cluster (incuding those immediately 
behind 
the cluster), and the lensing efficiency of red sequence nonmembers is $\beta=0.085$. It is 
therefore possible that the contaminating red sequence galaxies contain some lensing signal from 
background galaxies, but within the red sequence selection limits imposed here, this sample is only 
16\% of the red sequence galaxies. Therefore there is a fraction $\sim\!0.72\times0.16\simeq12\%$ 
of 
contaminating galaxies (with $\sim\!0.28\times0.16\simeq4\%$---those in the foreground---adding 
noise). The lensing signal in these galaxies is $\gamma_{+,\rm rs} \lesssim 
0.11\cdot\beta\cdot\gamma_+ = 0.11\cdot0.085\cdot0.10\approx9\times10^{-4}$, several times smaller 
than the statistical uncertainties (where $\gamma_+\approx0.1$ is a typical shear amplitude in the 
inner regions of galaxy clusters).

In summary, the red sequence gives a high-fidelity cluster member selection. It is important, 
however, to restrict this selection to the inner regions of clusters and to luminous galaxies (as 
shown in \cref{f:rspurity}), because the red sequence may contain some lensing signal. The purity 
of the red sequence as selected here is 84\%, so this contamination is not expected to be 
significant. Adding the red sequence members to the \Nmembers\ spectroscopically confirmed members 
gives a total of $N_m+N_{rs}=23,\!041$ members with an estimated contamination of 
$0.16\cdot N_{rs}/(N_m+N_{\rm rs})\approx8\%$.

\subsection{Photometric redshift contamination}

By taking a fixed width in velocity, we can simulate the members found by an accurate, unbiased 
photometric redshift criterion. The dotted lines in \cref{f:rspurity} show the fraction of galaxies 
that are within $\Delta z=0.03(1+z)$ (as expected for large ongoing photometric redshifts) but are 
not members of the cluster,\footnote{$\Delta z=0.03(1+z)$ corresponds to 
$\approx\!10,000\,\mathrm{km\,s^{-1}}$ at the median redshift of the sample, $z=0.15$.} as 
determined in \cref{s:specmembers}. The contamination is roughly independent of magnitude at all 
radii and at a level of $\sim13\%$ within 1 Mpc of the BCG, rising steeply beyond this radius. In 
terms of apparent magnitude the curves look similar in the range $m_r\lesssim23$, the range in 
which most of the selected red sequence galaxies are found. This contamination rises shallowly with 
redshift, reaching $\sim\!20\%$ at $z\gtrsim0.3$.

The radial dependence in \cref{f:rspurity} is shown in physical units instead of in units of 
$r_{200}$ because this is more generally used with photometric surveys where the physical size of 
each cluster is not known, and \cref{f:rspurity} gives an idea of the apertures that should be used 
to either search for clusters or characterize the cluster based on a red sequence sample.

Comparing the dotted and solid curves, it seems that there is not such a significant gain in using 
photo-$z$'s versus the red sequence. A photo-$z$ selection has the advantage that it selects a more 
representative population of the cluster, and that the red sequence depends on a single color (at 
least in this implementation) and it becomes less reliable when the 4000\AA\ break is not bracketed 
by the filters used. This is the case in our study for $z\gtrsim0.36$. It is also apparent, as with 
the red sequence, that a photo-$z$ selection becomes significantly contaminated beyond 
$r\sim1\,\mathrm{Mpc}$.

Finally, we note that the galaxies we refer to here (shown with the dotted lines) are not cluster 
members but also do not contribute a lensing signal, because they are too close behind. They are, 
indeed, likely to be part of the same large-scale structure of the cluster so would probably feel 
tidal torque from it similar to the actual member galaxies. Thus from the perspective of galaxy 
alignment measurements these galaxies should not dilute the signal significantly, nor introduce a 
lensing signal.

\subsection{Control Samples}\label{s:control}

We construct two catalogs as control samples to assess spurious contributions to our measurements. 
The shapes of objects in these two samples are unaffected by the cluster (and are mostly unrelated 
between objects in each sample), so their alignment signals (see \cref{s:ia}) should be consistent 
with zero. A departure from zero would mean that there is significant residual PSF ellipticity in 
the images, and therefore that the shape measurements are unreliable.

First, we use all stars in the magnitude range $17<m_r<22$, selected as outlined in 
\cref{s:photometry}, for a total of 443,321 stars. We choose the bright limit to avoid saturated 
stars, whereas the faint limit ensures that the star sample is not contaminated by faint, 
unresolved galaxies.

We also use all spectroscopically confirmed foreground galaxies, which are selected as all galaxies 
with peculiar velocities more negative than $-10,000\,\!\mathrm{km\,s^{-1}}$ in the rest-frame of 
the cluster. There are 3,666 spectroscopically confirmed foreground galaxies in the direction of 73 
clusters. The clusters with the most foreground galaxies are two of the highest-redshift clusters 
of the sample, namely MS~0451.6$-$0305 at $z=0.539$ and MACS~J0717.5+3745 at $z=0.544$, with 
$N_{\rm fg}=306$ and $N_{\rm fg}=304$, respectively.

\section{Measuring Intrinsic Alignments}\label{s:ia}

We measure the alignment signal of galaxies within clusters by weight-averaging the ellipticity 
components of all galaxies within a given radial annulus,
\begin{equation}\label{eq:avg}
 \langle\epsilon_i\rangle = \frac{\sum_n w_n \epsilon_{i,n}}{\sum_n w_n}\,,
\end{equation}
with weights equal to
\begin{equation}
 w_n = \frac1{\epsilon_{\rm int}^2 + \sigma_n^2}\,,
\end{equation}
where $\sigma_n$ is the measurement uncertainty on the ellipticity of the {\it n}-th galaxy. We 
assume an intrinsic (i.e., unlensed) galaxy ellipticity dispersion $\epsilon_{\rm 
int}=\sqrt{\langle\epsilon_i\epsilon_i\rangle}=0.25$. The uncertainty in \cref{eq:avg} is equal for 
both components and is given by $\sigma(\epsilon_i) = \left(\sum_nw_n\right)^{-1/2}$. In this work, 
we use the shapes of cluster members to measure three kinds of alignment: the alignment of 
(satellite) galaxies toward the center of the cluster, the alignment of galaxies with respect to 
the BCG orientation, and the alignment between satellite galaxies. These three quantities are 
detailed below.

Throughout, we refer to raw ellipticities as $e_i$, and to ellipticities that account for 
instrumental effects (i.e., PSF size in the case of gaussianized images) as $\epsilon_i$.

\subsection{Different alignment signals}

In this section we outline the different rotations we apply to the ellipticity measurements of 
\cref{s:shapes} in order to extract alignment signals within clusters.

\subsubsection{Satellite radial alignment}

We measure the alignment of galaxies with respect to the center of the cluster using ellipticity 
components rotated to a frame such that
\begin{align}\label{eq:epsilon}
 \epsilon_+ &= -(\epsilon_1\cos 2\theta + \epsilon_2\sin 2\theta) \\
 \epsilon_\times &= \epsilon_1\sin 2\theta - \epsilon_2\cos 2\theta\,
\end{align}
where $\epsilon_1$ and $\epsilon_2$ are the galaxy ellipticities in the cartesian frame, with 
$\epsilon_1$ measuring the ellipticity in the $x$ and $y$ directions, and $\epsilon_2$ in diagonal 
directions. Here $\theta$ is the azimuthal angle with respect to the center of the cluster. In 
this frame, $\epsilon_+$ measures the distortion in the tangential and radial directions while 
$\epsilon_\times$ measures the distortion at $\pm45^\circ$ from the radial direction \citep[see, 
e.g., Figure 1 of][for a diagram]{bernstein02}. Note that the definition of $\epsilon_+$ in 
\cref{eq:epsilon} has the opposite sign to that typically used in weak lensing analyses. For 
symmetry reasons the cross component, $\langle\epsilon_\times\rangle$, of an ensemble of clusters
should be consistent with zero (although a single cluster might have a preferred nonradial 
alignment direction such that $\langle\epsilon_\times\rangle\neq0$, the average over an ensemble of 
clusters must be zero), so it serves as a check for systematic effects. On the other hand, 
$\langle \epsilon_+\rangle<0$ indicates that galaxies are preferentially aligned in the tangential 
direction, as is the case for gravitationally lensed background galaxies, while 
$\langle\epsilon_+\rangle>0$ would indicate a radial alignment of the galaxies, which could be the 
case for cluster members. Finally, $\langle\epsilon_+\rangle=0$ implies that galaxies are 
randomly oriented toward the center of the cluster.

\subsubsection{Satellite-BCG alignment}

To measure the alignment between satellite galaxies and the BCG, we rotate the shapes and 
coordinates of satellites to a frame where the direction of $\epsilon_1>0$ coincides with the 
major axis of the BCG, namely
\begin{equation}\label{eq:bcg}
 \begin{split}
 \epsilon_1' &= \epsilon\cos\left[2\left(\phi-\phi_{\rm BCG}\right)\right] \\
 \epsilon_2' &= \epsilon\sin\left[2\left(\phi-\phi_{\rm BCG}\right)\right]\,,
 \end{split}
\end{equation}
where $\phi$ and $\phi_{\rm BCG}$ are the position angles of a satellite galaxy and the BCG, 
respectively, and $\epsilon\equiv(\epsilon_1^2+\epsilon_2^2)^{1/2}$ is the ellipticity of the 
galaxy. In this new frame, the BCG has ellipticity components $\epsilon_1'=\epsilon$ and 
$\epsilon_2'=0$. For the BCG position angles we use only GALFIT measurements (see \cref{s:galfit}), 
since these are expected to be more reliable for galaxies as large as BCGs. Analogous to the radial 
alignments, $\langle\epsilon_1'\rangle>0$ implies that satellite galaxies are oriented along the 
major axis of the BCG, $\langle\epsilon_1'\rangle<0$ that satellites are oriented along the BCG 
minor axis, and $\langle\epsilon_1'\rangle=0$ implies random orientations; $\epsilon_2'$ measures 
diagonal alignments so we expect $\langle\epsilon_2'\rangle=0$.

\subsubsection{Satellite-satellite alignment}\label{s:shapes-satsat}

Finally, we compute the alignment between satellite galaxies within clusters by calculating 
\cref{eq:epsilon} taking every satellite galaxy as a test galaxy (i.e., as the frame for $\theta$). 
BCGs are excluded from this analysis. This probes potential alignments of galaxies in substructures 
within the cluster. In principle, if there are $N$ members in a cluster, the number of pairs is 
equal to $N(N-1)/2$. However, we only use pairs for which a full circle can be averaged, to avoid 
averages that include mostly objects in the corners of the images where PSF residuals may be 
larger. This is a concern for massive, low redshift clusters, where $30\arcmin$ (half the side of 
the MegaCam image) is roughly equal to $r_{200}$ in the worst cases. To ensure that the average 
remains unbiased, therefore, we only include pairs such that the sum of the distance between the 
test galaxy and the center of the cluster and the separation between the two satellites is less 
than 90\% of the distance between the cluster center and the edge of the image.

\subsection{Shape Measurements}\label{s:shapes}

\begin{figure}
\centerline{\includegraphics[width=1.75in]{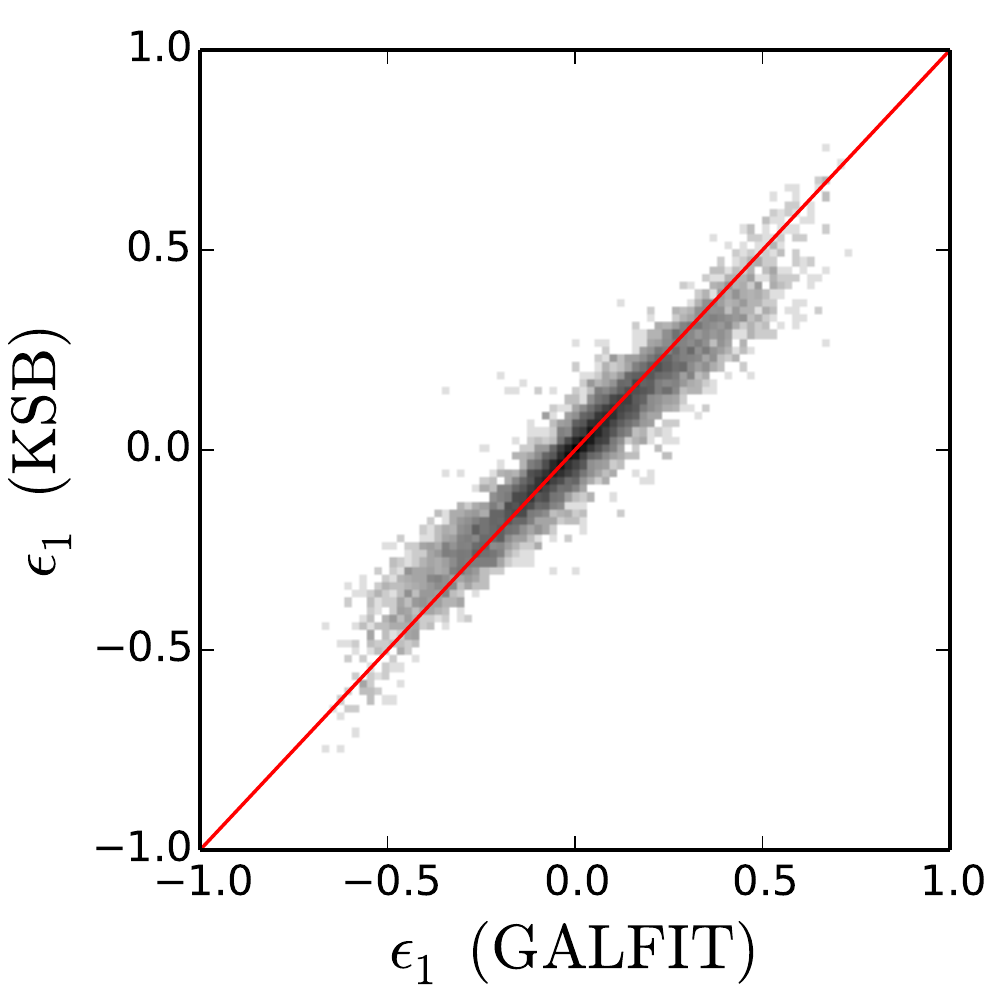}
            \includegraphics[width=1.75in]{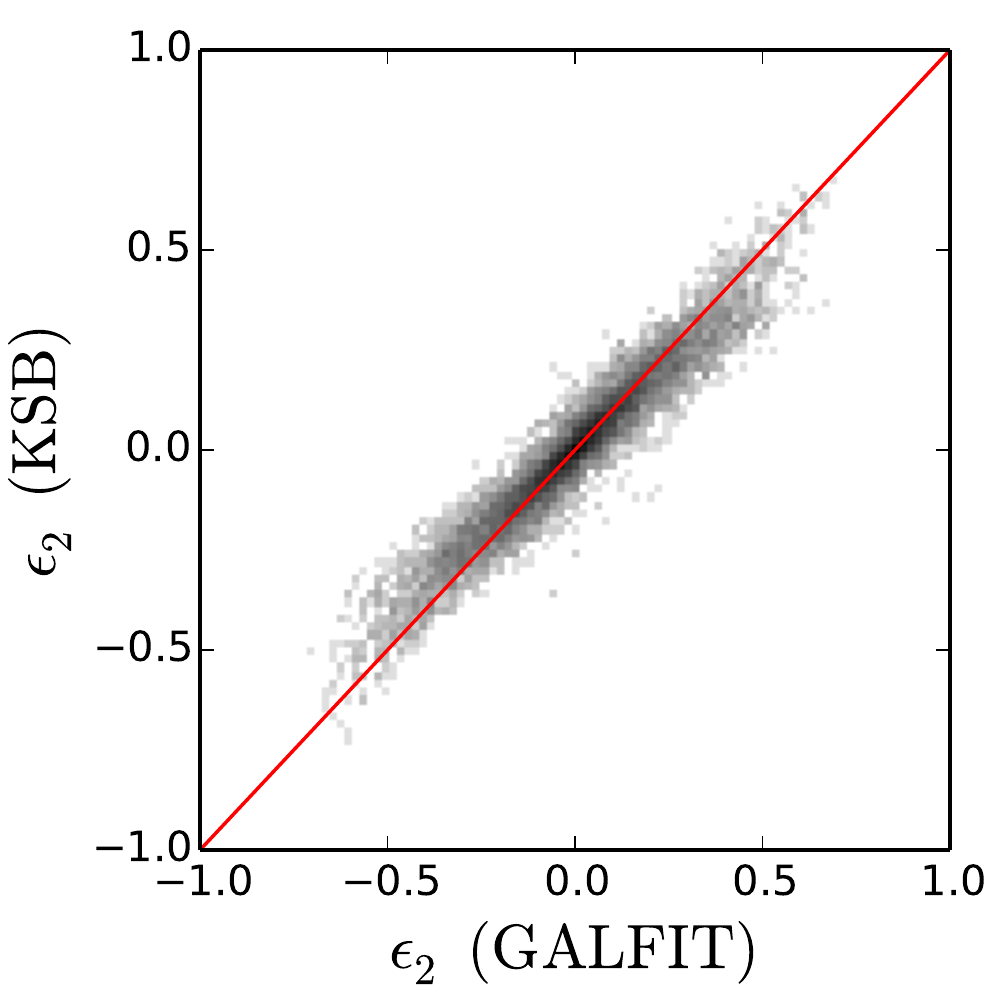}}
\centerline{\includegraphics[width=1.75in]{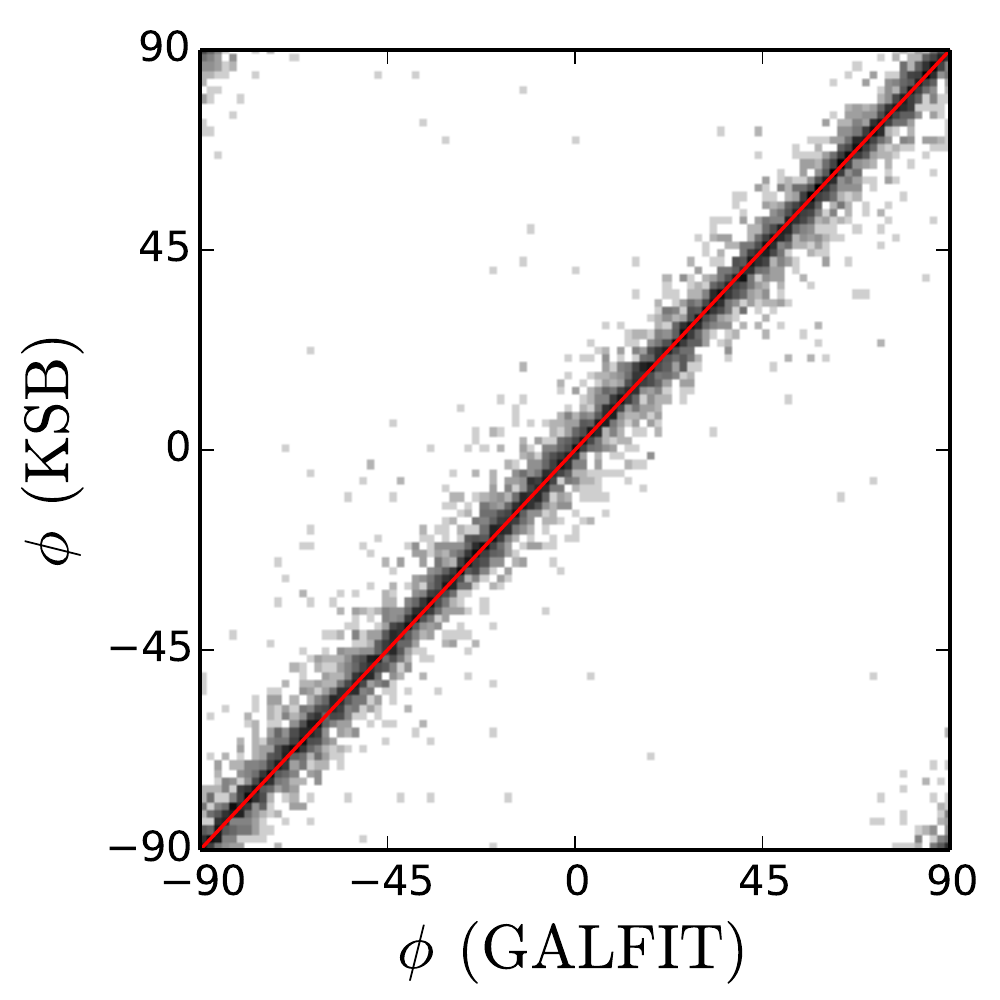}
            \includegraphics[width=1.75in]{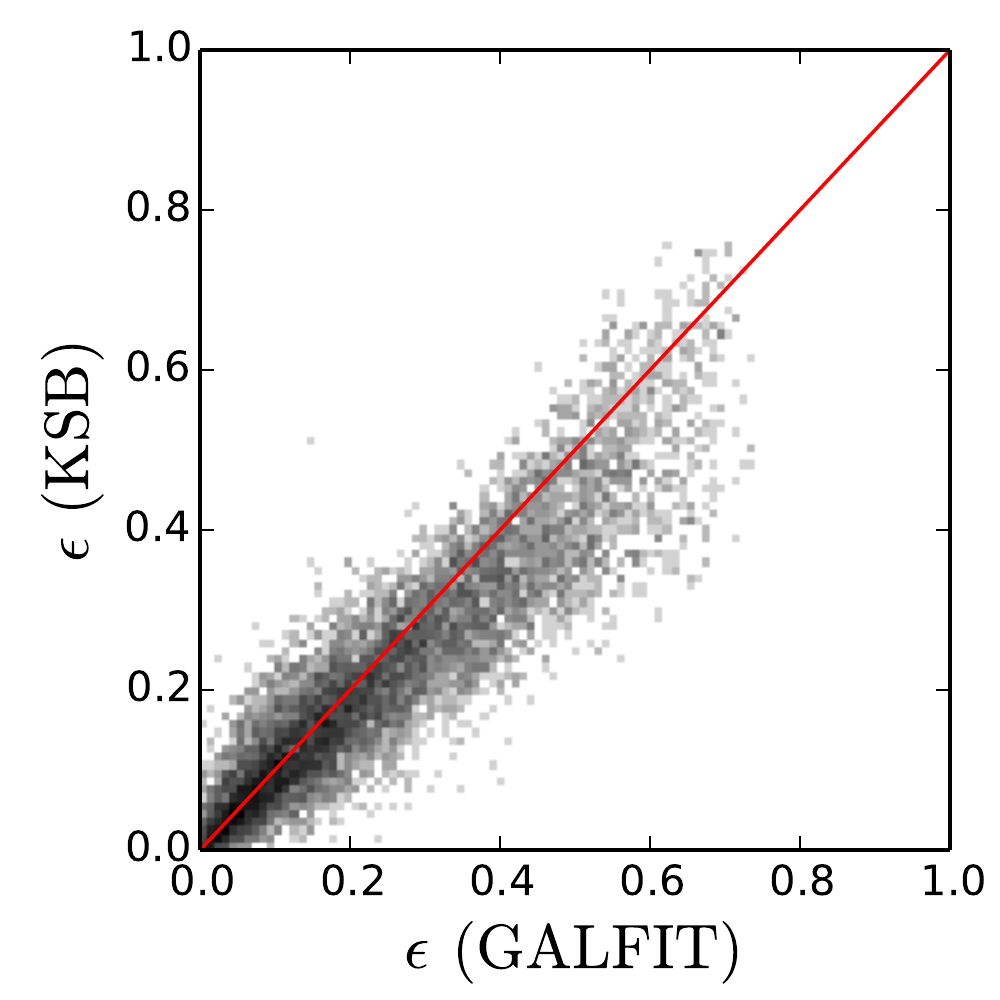}}
\caption{\small Comparison of shape measurements from KSB and GALFIT. Grey scales show the number 
of points per bin in logarithmic scale. Red lines show $y=x$. Top: Ellipticity components in 
cartesian coordinates. Bottom left: position angles, $\phi$, in degrees. The periodicity of $\phi$ 
(of 180 deg) can be seen in the top left and bottom right corners of the plot. Bottom right: 
galaxy ellipticities.}
\label{f:methods}
\end{figure}

\begin{figure*}
 \centerline{\includegraphics[width=3.4in]{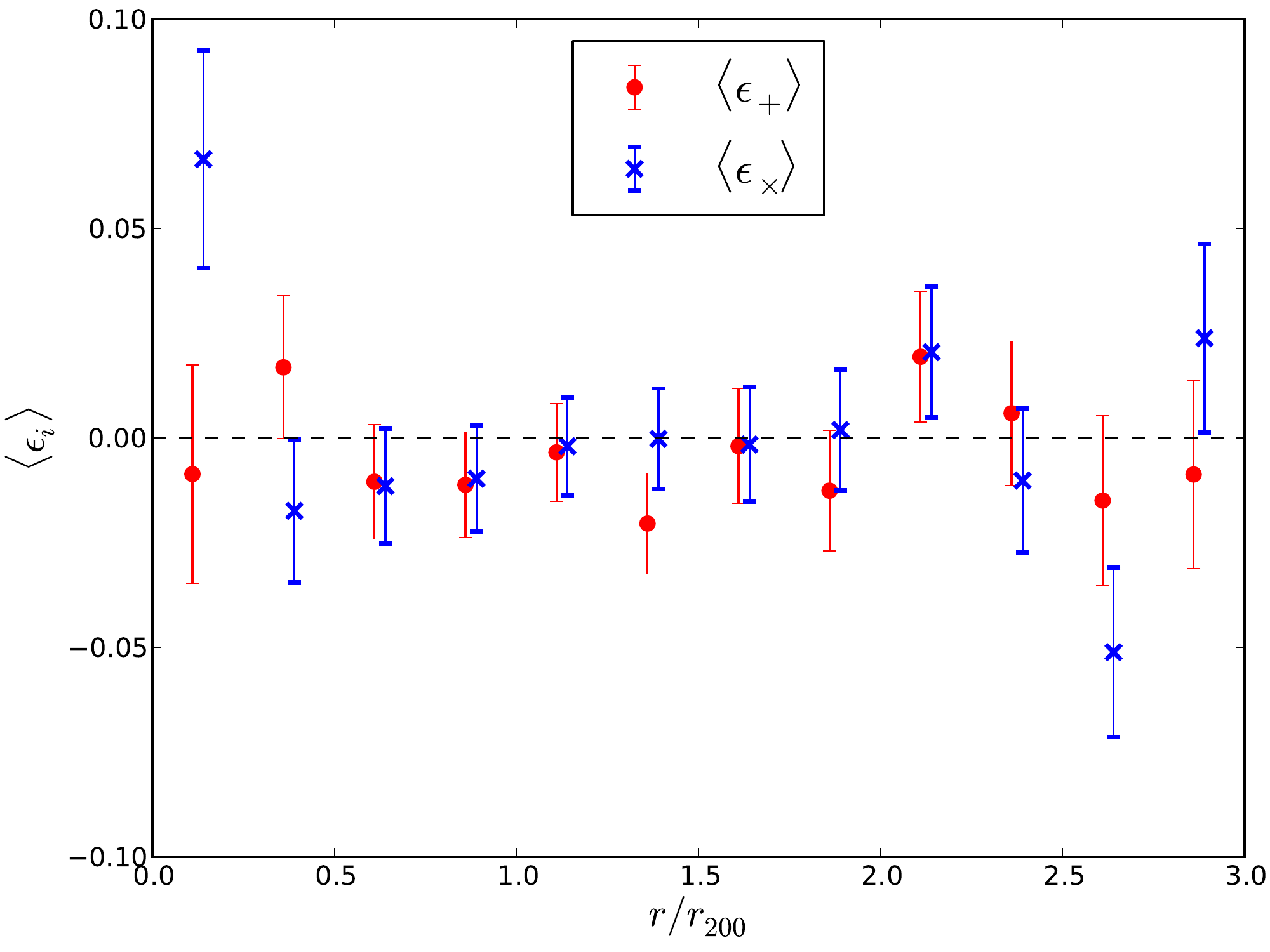}
             \includegraphics[width=3.5in]{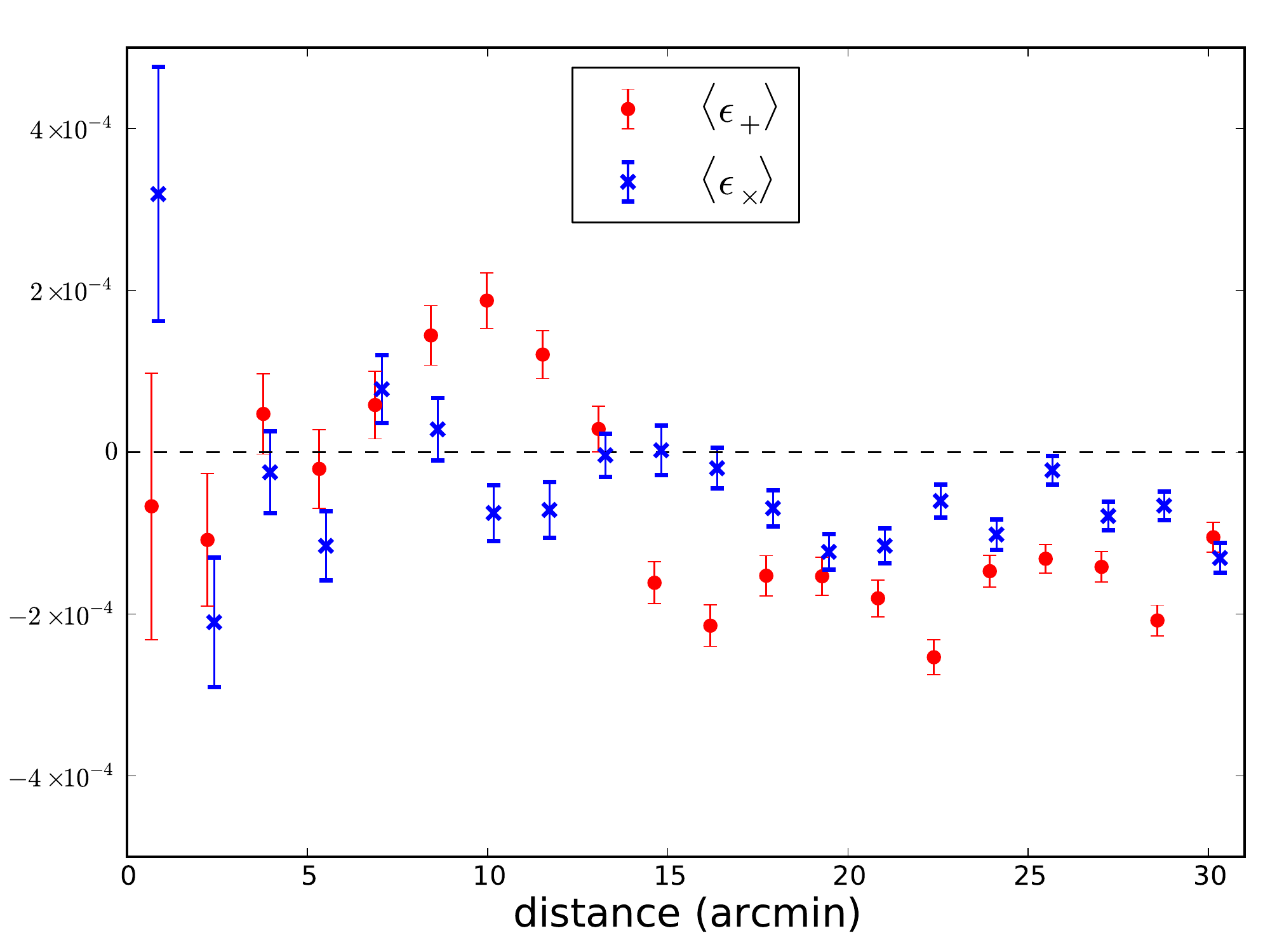}}
\caption{\small Alignment signal from control samples measured with KSB, with data points shifted 
horizontally for clarity. Left: 3,666 foreground galaxies in the direction of 73 clusters as a 
function of distance from the cluster, in units of \radius. Right: stars in the magnitude range 
$17<m_r<22$ as a function of angular distance from the center of each cluster. Typically, 
$r_{200}\sim10\arcmin$. Note the different vertical scales in each plot.}
\label{f:control}
\end{figure*}

Measuring galaxy shapes is a challenging endeavor, especially in the presence of noise and PSF 
anisotropies \citep[e.g.,][]{massey07,melchior12,kitching13}. For large (in units of the PSF), 
bright objects such as those used here, this should be less of a problem. Moreover, after 
gaussianization the PSF ellipticity is negligible. In this work we measure the shapes of member 
galaxies using two different methods, which allows us to test for consistency and robustness of the 
results. Below we give a brief outline of each method to highlight their differences; more details 
can be found in the original works.

Shapes are measured from the gaussianized images (\cref{s:photometry}). The PSF in these images is, 
by construction, circular, gaussian, and constant across the image. Therefore the shape measurement 
methods need to account for the blurring of the ellipticity by the PSF, but there are no systematic 
ellipticities in the images (to a high enough precision, see \cref{s:systematics}).

\subsubsection{Kaiser-Squires-Broadhurst (KSB)}

KSB was developed for weak lensing measurements by \cite{kaiser95} and revised by 
\cite{hoekstra98}. It measures shapes by estimating the central second moments $I_{ij}$ of the 
image 
fluxes to measure the two-component polarization
\begin{equation}
 e_1 = \frac{I_{11}-I_{22}}{I_{11}+I_{22}}\,;\,\,\,e_2 = \frac{2I_{12}}{I_{11}+I_{22}}\,.
\end{equation}
These measurents are weighted with a circular Gaussian of width $r_g$, which corresponds to the 
radius of maximum significance measured by KSB; this weight reduces shot noise in the measurements. 
Blurring by the PSF is corrected by the so-called pre-seeing shear polarizability, $P^\gamma$, 
which quantifies the effect of the convolution of the PSF to the image polarization, $e_i$ 
\citep{luppino97,hoekstra98}. The corrected ellipticity is then 
$\epsilon_i=e_i/P^\gamma$.\footnote{Because the PSF in our images has vanishing ellipticity by 
construction, the PSF correction of KSB is mathematically exact. This is not the case if the PSF is 
significantly elliptical.} Both $e_i$ and $P^\gamma$ are measured with the same radius, $r_g$, for 
each galaxy.

\subsubsection{GALFIT}\label{s:galfit}

GALFIT was developed by \cite{peng02}, having in mind the modeling of different components of 
galaxies for studies of galaxy structure and evolution. It attempts to model the light of a galaxy 
by fitting a multi-component generalized ellipse given by
\begin{equation}
 r = \left(\left\lvert x \right\rvert^{c+2} + 
           \left\lvert \frac{y}{q} \right\rvert^{c+2}\right)^{1/(c+2)}
\end{equation}
where a true ellipse has $c=0$, a boxy shape $c>0$ and a disky shape $c<0$; here $q$ is the 
minor-to-major axes ratio. Additionally, the position angle, $\phi$, is defined as the direction of 
the major axis. GALFIT accounts for the PSF model (in this case a single gaussian for each whole 
field) when measuring ellipticities. We use a simple \cite{sersic68} model for the surface 
brightness profile, $\ln I(r) \propto r^{1/n}$. Only galaxies with S\'ersic index $0.5<n<8$ and 
with axis ratio $q>0.15$ are included in the sample. We convert $q$ and $\phi$ to the same 
ellipticity measures of KSB through
\begin{equation}
 \epsilon_1 = \left(\frac{1-q}{1+q}\right)\cos2\phi \,;\,\,\,
 \epsilon_2 = \left(\frac{1-q}{1+q}\right)\sin2\phi\,.
\end{equation}

\subsection{Systematic effects}\label{s:systematics}

Because weak lensing measurements rely on averages of a large number of small signals, they are 
more prone to systematic effects than photometry and require more agressive masks. Therefore some 
spectroscopic members (all of which are in our photometric catalog) are not included in the shape 
catalogs. Moreover, the KSB and GALFIT catalogs are not the same since both have different 
requirements on, e.g., the size of an object and blending with nearby objects to estimate a 
reliable shape. Of the \Nmembers\ spectroscopic members, 13,966 have a KSB shape measurement and 
13,360 have a GALFIT measurement, with an overlap of 12,160 galaxies and a total of 14,250 
galaxies with a shape measurement. Similarly, of 23,041 spectroscopic+RS members, 20,493 have KSB 
measurements and 18,511 have GALFIT measurements. The smaller number of objects with GALFIT 
measurements comes mainly from high-redshift galaxies (compare \Cref{t:shear-spec,t:shear-rs}). 
This is because small, faint galaxies are harder for GALFIT to fit, while KSB is well-suited to 
measure the shapes of faint (background) galaxies.

We only consider galaxies with shape measurements from either method in this work, except 
for the assessment of the red sequence in \cref{s:photmembers}. \cref{f:methods} compares the shape 
parameters for all spectroscopic members that have valid KSB and GALFIT estimates. While the 
measurements generally agree, there is a small but noticeable difference for large-ellipticity 
objects, such that KSB estimates lower ellipticities than GALFIT. This effect is present with more 
or less the same magnitude for all clusters; it is a genuine difference between the two methods 
(for our particular dataset), and there is an indication that this effect may be more pronounced 
for smaller objects. This difference is due to higher-order corrections that are not implemented in 
KSB, which become important at large ellipticities \citep{viola11}. As we show in \cref{s:results} 
this has no impact on our results, so we do not explore this issue further.

As a further test, \cref{f:control} shows the alignment signals of the control samples. As 
expected, foreground galaxies have a signal consistent with zero in both ellipticity components at 
all radii, with large errorbars due to small statistics. The average ellipticities of stars are 
different from zero at significant levels in most of the radial range. However, the average 
ellipticity is constrained to $\langle\epsilon_i\rangle\lesssim2\times10^{-4}$ at all radii, an 
order of magnitude smaller than the statistical errors in the alignments of cluster members. Thus 
any systematic effects arising from PSF uncertainty or other instrumental biases are controlled to 
much lower values than the statistical uncertainties, and can be neglected for the purposes of this 
work.

Finally, the gaussianization of the images makes the PSF round and homogeneous across an image but 
produces anisotropic (correlated) noise, which could introduce noise bias in our measurements. The 
level of anisotropy can be assessed by measuring star ellipticities as a function of magnitude: 
if noise is highly anisotropic then noisier measurements would show, on average, a larger 
anisotropy than high-S/N measurements. We test this by comparing the ellipticities of stars as a 
function of magnitude (for $18\leq m_r\leq22$), and find that the average ellipticities are 
consistent with the levels shown in \cref{f:control}. Moreover, we use galaxies whose number 
density 
drops rapidly beyond $m_r\sim18$, and are typically 8 times larger than the PSF. We conclude that 
anisotropic noise can be safely neglected in this study.

\section{Results}\label{s:results}

In this section we present and discuss the main results of this paper. We refer to \cref{s:ia} 
for details on the calculations that lead to the values reported here and a discussion of 
systematic effects.

\begin{table*}
\begin{center}
\caption{Average ellipticity components of spectroscopic members.}
\label{t:shear-spec}
\begin{tabular}{l | r r r c | r r r c}
\hline\hline
 &  & \multicolumn{2}{c}{KSB} &  &  & \multicolumn{2}{c}{GALFIT} &  \\
\hline
Sample & $N_{\rm gal}$\tma & \multicolumn{1}{c}{$\langle \epsilon_+ \rangle$} & 
\multicolumn{1}{c}{$\langle \epsilon_\times \rangle$} & 
$\sigma(\epsilon_i)$\tmb & $N_{\rm gal}$\tma & \multicolumn{1}{c}{$\langle 
\epsilon_+ \rangle$} & 
\multicolumn{1}{c}{$\langle \epsilon_\times \rangle$} & 
\multicolumn{1}{c}{$\sigma(\epsilon_i)$\tmb} \\[0.5ex]
\hline
All & 8,510 & $-$0.0037 & $-$0.0014 & 0.0027 & 8,014 & 0.0004 & $-$0.0009 & 0.0031 \\
[0.8ex]
$z<0.14$ & 4,170 & 0.0002 & $-$0.0000 & 0.0039 & 4,612 & 0.0029 & $-$0.0003 & 0.0038 \\
$z\geq0.14$ & 4,340 & $-$0.0074 & $-$0.0027 & 0.0038 & 3,402 & $-$0.0042 & $-$0.0020 & 0.0053 \\
[0.8ex]
$M_{200}<7\times10^{14}M_\odot$ & 2,287 & $-$0.0059 & 0.0051 & 0.0052 & 2,277 & $-$0.0057 & 0.0041 
& 0.0057 \\
$M_{200}\geq7\times10^{14}M_\odot$ & 6,223 & $-$0.0029 & $-$0.0038 & 0.0032 & 5,737 & 0.0030 & 
$-$0.0030 & 0.0037 \\
[0.8ex]
Relaxed & 3,233 & $-$0.0037 & $-$0.0022 & 0.0044 & 3,058 & $-$0.0038 & $-$0.0025 & 0.0050 \\
Disturbed & 5,277 & $-$0.0036 & $-$0.0009 & 0.0034 & 4,956 & 0.0031 & 0.0001 & 0.0040 \\
[0.8ex]
$M_r\leq-21$ & 4,101 & $-$0.0031 & $-$0.0016 & 0.0039 & 3,922 & 0.0009 & $-$-0.0000 & 0.0044 \\
$M_r>-21$ & 4,409 & $-$0.0042 & $-$0.0012 & 0.0038 & 4,092 & $-$0.0001 & $-$0.0018 & 0.0044 \\
[0.8ex]
RS & 5,806 & $-$0.0008 & $-$0.0009 & 0.0033 & 5,595 & 0.0010 & $-$0.0001 & 0.0037 \\
Non-RS & 2,704 & $-$0.0099 & $-$0.0025 & 0.0048 & 2,419 & $-$0.0012 & $-$0.0031 & 0.0059 \\
\hline
\end{tabular}
\end{center}
\tablefoottext{a}{Number of galaxies used for the average, within $r_{200}$.}
\tablefoottext{b}{68\% confidence measurement uncertainties on the average ellipticities.}
\end{table*}

\begin{table*}
\begin{center}
\caption{Average ellipticity components of spectroscopic plus red sequence members.}
\label{t:shear-rs}
\begin{tabular}{l | r r r c | r r r c}
\hline\hline
 &  & \multicolumn{2}{c}{KSB} &  &  & \multicolumn{2}{c}{GALFIT} &  \\
\hline
Sample & $N_{\rm gal}$ & \multicolumn{1}{c}{$\langle \epsilon_+ \rangle$} & 
\multicolumn{1}{c}{$\langle \epsilon_\times \rangle$} & 
$\sigma(\epsilon_i)$ & $N_{\rm gal}$ & \multicolumn{1}{c}{$\langle 
\epsilon_+ \rangle$} & 
\multicolumn{1}{c}{$\langle \epsilon_\times \rangle$} & 
\multicolumn{1}{c}{$\sigma(\epsilon_i)$} \\[0.5ex]
\hline
All & 15,905 & $-$0.0022 & $-$0.0021 & 0.0020 & 12,930 & 0.0000 & $-$0.0008 & 0.0026 \\
[0.8ex]
$z<0.14$ & 6,407 & $-$0.0020 & $-$0.0019 & 0.0031 & 6,233 & 0.0013 & 0.0002 & 0.0033 \\
$z\geq0.14$ & 9,498 & $-$0.0023 & $-$0.0022 & 0.0026 & 6,697 & $-$0.0019 & $-$0.0023 & 0.0041 \\
[0.8ex]
$M_{200}<7\times10^{14}M_\odot$ & 5,394 & $-$0.0039 & $-$0.0010 & 0.0034 & 4,345 & $-$0.0057 & 
0.0009 & 0.0044 \\
$M_{200}\geq7\times10^{14}M_\odot$ & 10,511 & $-$0.0013 & $-$0.0027 & 0.0024 & 8,585 & 0.0031 & 
$-$0.0017 & 0.0032 \\
[0.8ex]
Relaxed & 7,504 & $-$0.0034 & $-$0.0051 & 0.0029 & 5,903 & $-$0.0044 & $-$0.0038 & 0.0038 \\
Disturbed & 8,401 & $-$0.0011 & 0.0006 & 0.0027 & 7,027 & 0.0039 & 0.0018 & 0.0035 \\
[0.8ex]
$M_r\leq-21$ & 5,394 & $-$0.0020 & $-$0.0011 & 0.0034 & 4,912 & 0.0008 & 0.0004 & 0.0040 \\
$M_r>-21$ & 10,511 & $-$0.0023 & $-$0.0026 & 0.0024 & 8,018 & $-$0.0005 & $-$0.0017 & 0.0034 \\
\hline
\end{tabular}
\end{center}
\end{table*}

\begin{figure}
 \centerline{\includegraphics[width=3.5in]{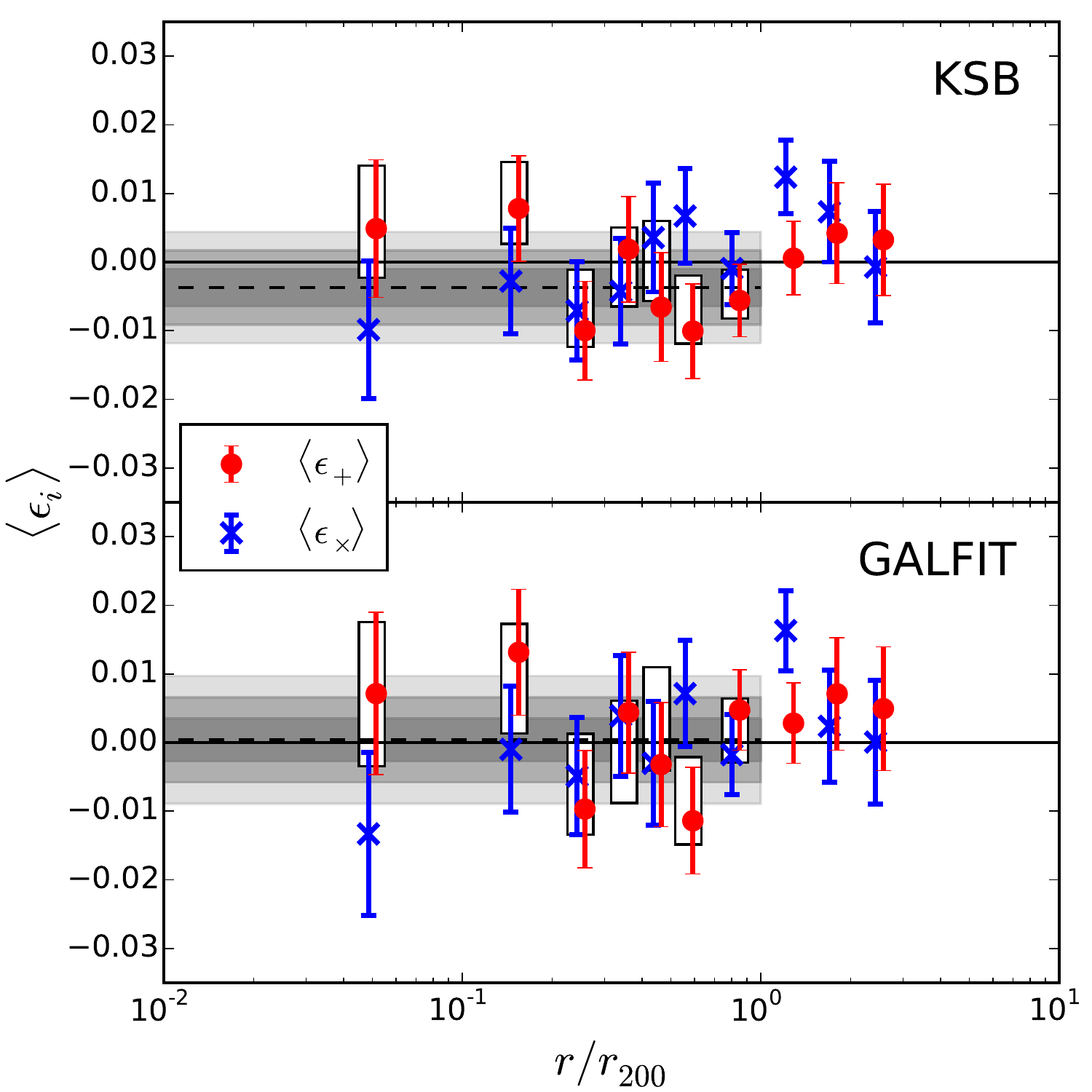}}
\caption{Average alignment of all spectroscopically confirmed members out to $3r_{200}$. The 
\textit{top} panel shows the results from KSB while the \textit{bottom} panel shows those from 
GALFIT. Shaded bands show the 1, 2 and 3$\sigma$ uncertainties in the overall average and white 
bars show the $1\sigma$ range for $\langle\epsilon_+\rangle$ from the enhanced sample including red 
sequence members. Points are slightly shifted horizontally for clarity.}
\label{f:shear-all}
\end{figure}

\subsection{Satellite radial alignment}\label{s:radial}

\begin{figure}
 \centerline{\includegraphics[width=3.5in]{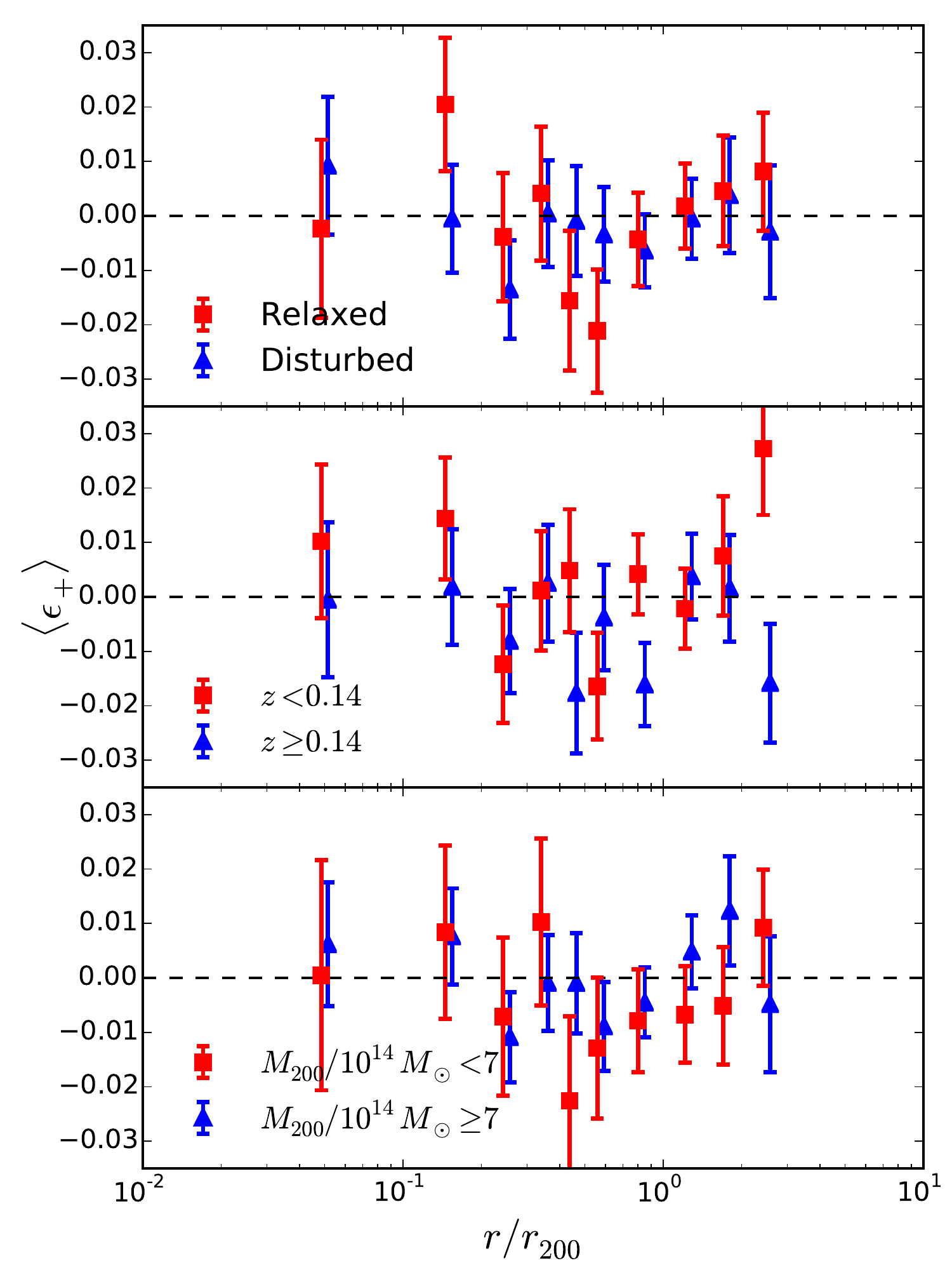}}
\caption{\small Average alignment $\langle \epsilon_+ \rangle$ from KSB for spectroscopically 
confirmed members, divided by cluster properties: by dynamical state (\textit{top}; see 
\cref{s:substructure}), redshift (\textit{middle}), and dynamical mass (\textit{bottom}). The 
latter two thresholds are the median values of the cluster sample.}
\label{f:shear-clsub}
\end{figure}

\cref{f:shear-all} shows the average radial alignment for all spectroscopically confirmed cluster 
members with good ellipticity measurements from KSB and GALFIT in annuli around the cluster center. 
Both methods show that the intrinsic alignment signal of cluster members is consistent with zero 
across all radii. Hereafter, we choose to quote average values within $r_{200}$ since, strictly 
speaking, this is the input required by the halo model (see \cref{s:ps}). Within \radius, the 
alignment of spectroscopic members is constrained to an average of 
$\langle\epsilon_+\rangle=-0.0037\pm0.0027$ with KSB and $\langle\epsilon_+\rangle=0.0004\pm0.0031$ 
with GALFIT at 68\% confidence. The cross components are also consistent with zero. Including red 
sequence members roughly doubles the number of galaxies used and confirms the latter result, with 
$\langle\epsilon_+\rangle=-0.0022\pm0.0020$ and $\langle\epsilon_+\rangle=0.0000\pm0.0026$ with 
KSB and GALFIT, respectively.

Our results are consistent with the nondetection of satellite radial alignments in massive clusters 
at $z>0.5$ \citep{hung12}, based on $\sim\!500$ spectroscopic members in the inner $\sim\!500$ kpc 
of clusters, using imaging from the Hubble Space Telescope (HST), and also with measurements at 
smaller masses from  photometrically-selected galaxy groups from SDSS \citep{hao11,chisari14} and 
spectroscopically-selected galaxy groups from the Galaxy And Mass Assembly (GAMA) survey 
\citep{schneider13}. Our results suggest that the stars in galaxies within clusters do not feel a 
strong enough tidal torque to be aligned toward the center of the cluster, in contrast with results 
from simulations which find strong alignments even when accounting for differences in the response 
between stars and dark matter which naturally occurs in hydrodynamical simulations 
\citep[][Velliscig et al.\ in prep]{pereira10,tenneti14}. An obvious consideration from the 
observational point of view is miscentering: whether the chosen cluster center is really the 
minimum of the cluster potential. This effect can be measured statistically with stacked weak 
lensing measurements \citep[e.g.,][]{george12} but is otherwise hard to assess observationally. At 
least in very relaxed clusters, BCGs are typically very close to the peak of the gas distribution 
\citep[e.g.,][]{lin04,mahdavi13}, which is closely matched to the dark matter distribution 
\citep{faltenbacher07b}. We can therefore test, to some extent, whether miscentering could be 
diluting an alignment signal by isolating relaxed clusters as discussed in \cref{s:substructure}. 
However, as shown in the top panel of \cref{f:shear-clsub}, we do not detect any alignment signal 
neither from relaxed nor from disturbed clusters. Thus we conclude that our results are robust to 
miscentering effects and that, statistically, satellite galaxies do not align toward the centers of 
clusters.

As discussed by \cite{hao11}, the redshift evolution of satellite radial alignments, if any, 
contains valuable information as to whether these alignments are produced during the formation of 
clusters or an evolving product of tidal torques within clusters. The middle panel of 
\cref{f:shear-clsub} shows that the alignment signal is consistent with zero across redshift, 
suggesting that neither of these processes is sufficient to sustain radial alignments over 
cosmological time. Furthermore, the bottom panel of \cref{f:shear-clsub} shows that this 
nondetection is also independent of cluster mass. We further tested whether any orientation bias, 
in the sense that we might have clusters viewed preferentially along their major axis, could have 
any effects on our results. To do this, we divided the cluster sample by BCG elongation, assuming 
that BCGs that look rounder might actually be elongated along the line-of-sight. Both cluster 
samples have radial alignments consistent with zero (not shown), arguing that a possible 
orientation bias is not a problem here.

\begin{figure}
 \centerline{\includegraphics[width=3.5in]{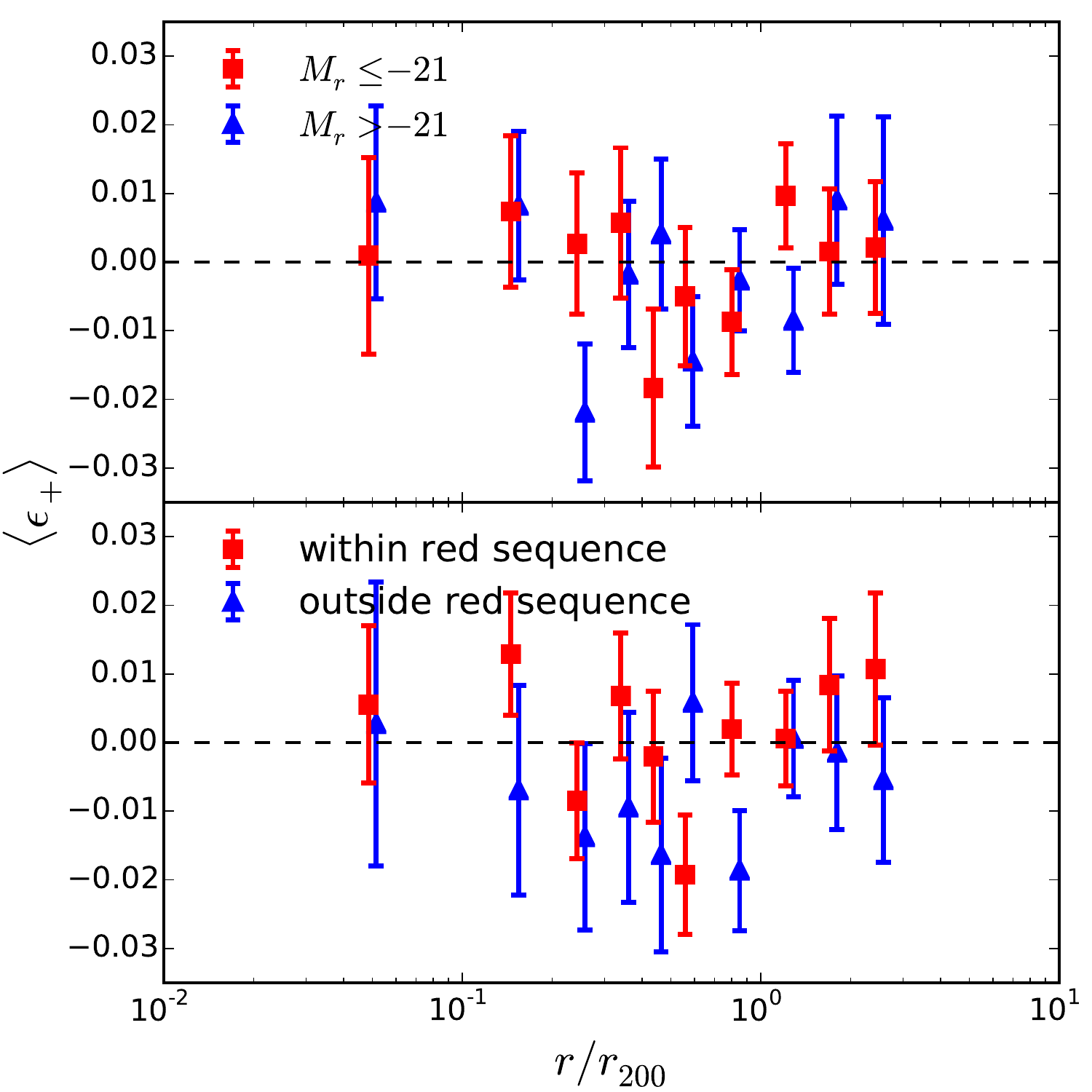}}
\caption{\small Average alignment $\langle \epsilon_+ \rangle$ from KSB for spectroscopically 
confirmed members divided by galaxy properties: by rest-frame $r$-band absolute magnitude 
(\textit{top}), and color with respect to each cluster's red sequence (\textit{bottom}).}
\label{f:shear-galsub}
\end{figure}

In any of the two scenarios mentioned above (namely tidal and primordial alignments), radial 
alignments could show a different pattern for galaxies with different formation histories. We 
investigate this by splitting the galaxy sample by galaxy luminosity (as a proxy for galaxy mass) 
and color---since bluer galaxies have been accreted more recently. To split by galaxy color we use 
each cluster's red sequence, which depends linearly on apparent magnitude, as outlined in 
\cref{s:photmembers}. As seen in \cref{f:shear-galsub}, we find no radial alignments consistently 
across galaxy colors and luminosities.

The results discussed above are summarized in \Cref{t:shear-spec,t:shear-rs} for spectroscopic and 
spectroscopic plus red sequence member samples, respectively.

\subsection{Satellite-BCG alignment}\label{s:bcgalignment}

The second type of alignment we explore is the alignment of the satellite orientations with the BCG 
orientation (cf.\ \cref{eq:bcg}). A large number of observations suggest that BCGs are on 
average oriented along the major axes of clusters themselves 
\citep[e.g.,][]{sastry68,binggeli82,faltenbacher07,niederste10,hao11}, and there is evidence that 
the velocity dispersion of satellite galaxies is typically larger along the BCG major axis 
\citep{skielboe12}. It is possible, then, that the BCG orientation represents a preferred infall 
direction. If this is the case, it is possible that galaxies would be aligned toward this infall 
direction.

\begin{figure}
 \centerline{\includegraphics[width=3.5in]{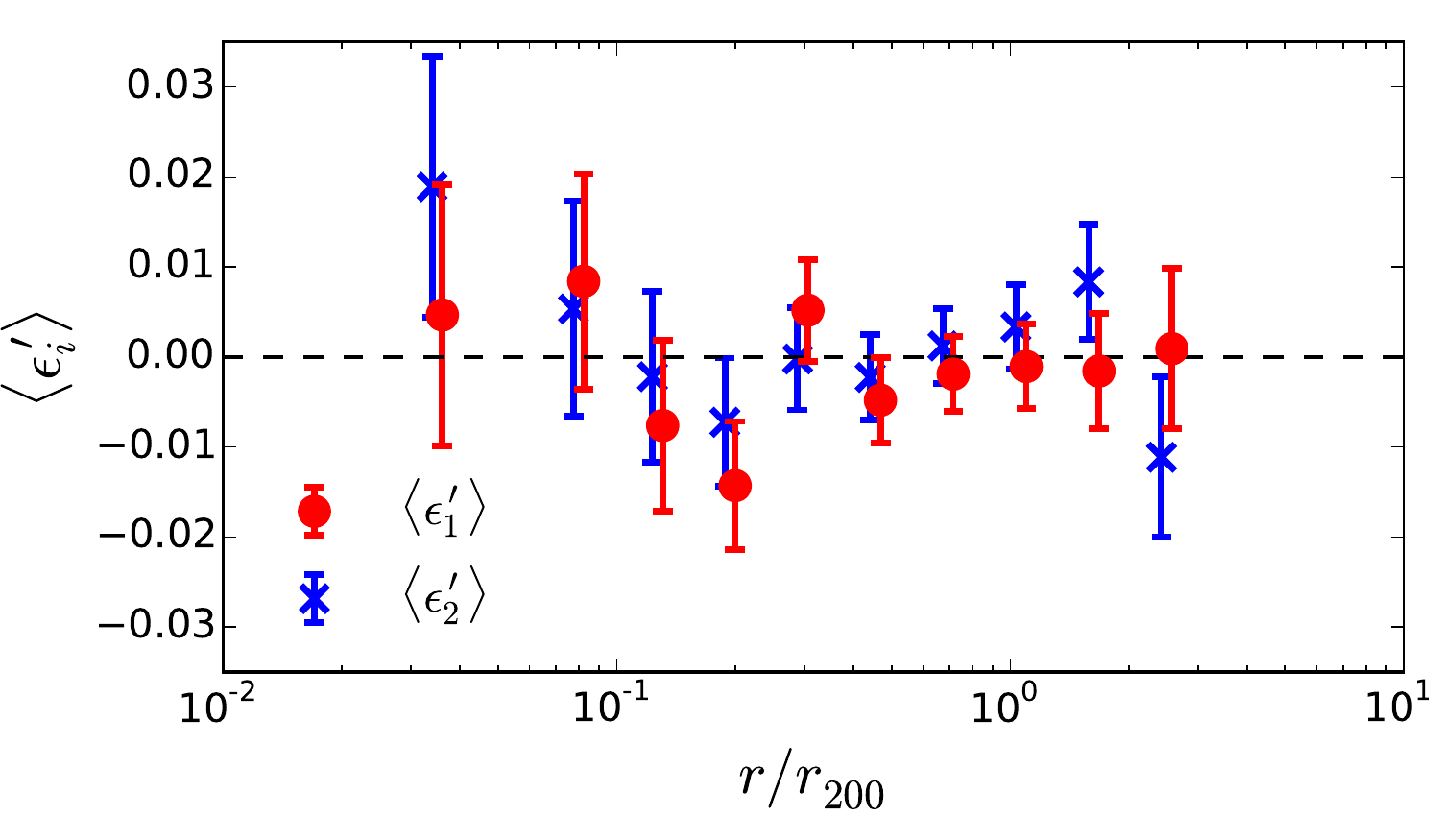}}
\caption{\small Mean ellipticity components of spectroscopic plus red sequence satellite galaxies 
in a frame rotated by the position angle of the BCG, probing the alignment of satellites with the 
cluster BCG. BCG position angles are measured with GALFIT, while the shapes of satellite galaxies 
are measured with KSB. Red circles show alignments with respect to the major ($\epsilon_1'>0$) and 
minor ($\epsilon_1'<0$) axes of the BCG, while blue crosses show alignments at $45^\circ$ 
rotations.}
\label{f:bcg}
\end{figure}

\cref{f:bcg} shows the alignment of galaxies with the major axis of the BCGs measured with KSB as a 
function of radius, for the full sample of spectroscopic plus red sequence members. As in the case 
of radial alignments, the data are also consistent with no satellite-BCG alignments at all 
distances. The average KSB signal within $r_{200}$ is $\langle\epsilon_1'\rangle=-0.0021\pm0.0022$; 
the average GALFIT signal is $\langle\epsilon_1'\rangle=-0.0024\pm0.0029$. We also split the sample 
as in the preceeding section, and find no signal for all galaxy and cluster subsamples. As a 
consistency check, we also find that the distribution of position angles, $\lvert\phi-\phi_{\rm 
BCG}\rvert$, is consistent with a random distribution.

Finally, we averaged not in annular bins but in cartesian coordinates $\{x,y\}$, to check if the 
satellite-BCG alignment could be happening only along a preferential direction, such that the 
azimuthal average would dilute the signal. We also found a null signal in this case (not shown).

\subsection{Satellite-satellite alignment}\label{s:satsat}

We have shown in \cref{s:radial} that satellite galaxies are not aligned toward the centers 
of clusters. If galaxies reside within substructures themselves, then these substructures might 
have tidally aligned galaxies toward them. If the tidal torque of the cluster is not enough to 
overcome these substructure-scale alignments, then maybe we can observe an alignment signal at 
small separations, between satellite galaxies. After excluding data near the edges of the images 
(see \cref{s:shapes-satsat}), we use a total of $3.93\times10^6$ satellite pairs. \cref{f:satsat} 
shows the alignment signal between satellites averaged over all clusters, as a function of distance 
between satellites, for the full spectroscopic plus red sequence member sample. In this case we 
split the sample into two radial bins, namely (test) galaxies within and outside $0.25r_{200}$, 
which corresponds to the scale radius of a cluster with a concentration $c_{200}=4$ \citep[roughly 
what is expected for massive clusters; e.g.,][]{duffy08}, but the results are similar when 
splitting the sample at other radii.

The leftmost bins in \cref{f:satsat} show the signal from substructure: outer bins probe the 
radial alignment between galaxies at large distances. It might be expected that substructure in the 
outskirts of clusters would contain an alignment signal since, presumably, they have been accreted 
more recently. As in the preceeding sections, we do not observe any alignment signal for the full 
cluster sample, nor for relaxed or disturbed clusters, at any radii. We note however that the last 
data point in \cref{f:satsat} is significantly nonzero, but so is the cross component. This 
suggests that at these distances measurements are affected by systematic effects, mainly because a 
large fraction of the pairs consist of two galaxies at the edges of the images. Moreover, this 
data point shows the alignments between satellites at opposite sides of the cluster; i.e., it is 
not a measurement of alignment within cluster substructure.

Since we do not detect any alignment signal for clusters at different redshifts and at different 
dynamical stages, we conclude that tidal torques in clusters, or in substructures within them, do 
not result in significant alignments of the stellar content of galaxies at any scale (neither 
toward the center nor between galaxies). It may be possible to bring this in line with the strong 
alignments measured in $N$-body simulations by invoking a misalignment between the stellar and dark 
matter distributions \citep[e.g.,][]{okumura09,tenneti14}. However, such analysis is beyond the 
scope of this paper.

\subsection{Is there an agreement on the level of galaxy alignments in groups and clusters?}
\label{s:agreement}

As discussed above, previous studies have reported various levels of alignment of satellite 
galaxies in clusters using different estimators. We expect such a lack of agreement to arise for 
two main reasons: the quality of the images used to measure galaxy shape parameters, and the use of 
shape measurements that are prone to systematic effects, e.g., isophotal measurements. The latter 
effect was studied by \cite{hao11} in detail; they found significant radial alignments only when 
using isophotal shape measurements, and that the strength of these alignments depends on apparent 
magnitude but \textit{not} on absolute magnitude, a strong suggestion that the detection is an 
artifact. Specifically, isophotal measurements are subject to severe contamination from the BCG, 
which can extend over a few hundred kpc in the case of massive clusters. As to the first cause, the 
quality of imaging data used by different groups varies significantly. To our knowledge, 
\cite{plionis03} were the first to use CCD photometry to measure galaxy alignments. They found a 
significant anisotropy in the (isophotal) position angles of satellite galaxies of Abell 521 
(though 
they used photographic plates for their statistical study of alignments in clusters). There are 
also 
recent studies, however, who used position angle measurements extracted from scanned photographic 
plates \citep[e.g.,][]{baier03,panko09,godlowski10}, both of which are of noticeably lower 
quality than present-day observations. Moreover, these works typically used single-band information 
to select cluster members, yielding an unknown (and likely low) sample purity.

Most recent studies have used data from SDSS because of its unmatched statistical power. 
These data, while of very high quality compared to photographic plate measurements, are several 
magnitudes shallower than our MegaCam data and taken under less ideal conditions (with seeing a 
factor 2 larger). Conversely, \cite{hung12} have used deep, high-quality HST imaging to measure 
galaxy alignments, finding no evidence for galaxy alignments within clusters. As in our analysis, 
\cite{hung12} have considered spectroscopically-confirmed cluster members, thus in addition to the 
superior photometry, both works have a cleaner member sample, which is key to the interpretation of 
the signal. \cite{schneider13} also used a sample of spectroscopically-confirmed group members, 
plus a shape measurement method that was specifically calibrated to weak lensing measurements 
\citep{mandelbaum05}, and found no significant evidence for alignments. Finally, \cite{chisari14} 
measured galaxy alignments in photometrically-selected galaxy groups and clusters in SDSS Stripe 
82, fully accounting for photometric redshift uncertainties, and constrain alignments to similar 
values as those found here.

The fact that all recent measurements that use high-quality imaging and properly calibrated 
shape measurements have yielded null detections \citep[][plus the present 
study]{hao11,hung12,schneider13,chisari14} leads us to conclude that there is no evidence for 
intrinsic alignments of satellite galaxies in galaxy groups or clusters to the level of 
uncertainty achievable with current datasets (both statistical and systematic).

\section{Contamination to Cosmic Shear Measurements}\label{s:ps}

In this section we explore the impact that the measured galaxy alignments in clusters can have on 
future cosmic shear measurements. We quantify the contribution of intrinsic alignments to cosmic 
shear measurements through the matter and intrinsic alignment power spectra, which can be defined as
\begin{equation}\label{eq:EBspectraDef}
\begin{split}
 \left\langle\tilde{\gamma}^{I*} (\kv)\densshear(\kv')\right\rangle &=
(2\pi)^3\delta^{(3)}_{D}\left(\kv-\kv'\right) P_{II}(\kv) \\
 \left\langle \delta^{*}(\kv)\densshear(\kv')\right\rangle &=
(2\pi)^3\delta^{(3)}_{D}\left(\kv-\kv'\right) P_{GI}(\kv)\,.
\end{split}
\end{equation}
Here, $\densshear=(1+\delta_g)\gamma^I$ is the (projected) ellipticity field weighted by the 
galaxy density, $\delta_g$, and $P_{II}(\kv)$ and $P_{GI}(\kv)$ are the II and GI contributions to 
the power spectrum including a prescription for nonlinear evolution \citep[i.e, nonlinear power 
spectra, see][]{smith03,bridle07}, respectively; $\delta^{*}$ is the complex conjugate of the 
Fourier transform of the matter density contrast, $\delta(\rv) = (\rho(\rv)-{\bar \rho})/{\bar 
\rho}$ is the matter overdensity with respect to the average density of the Universe, 
$\tilde{\gamma}^{I*}$ indicates the complex conjugate of $\tilde\gamma^I$, and $\delta_D$ is a 
Dirac delta function.

Additionally, we translate the 3-dimensional power spectra discussed above into (observable) 
angular power spectra, $C_\ell$, using the \cite{limber53} approximation \cite[e.g.,][]{kaiser92}. 
We use a source redshift distribution given by
\begin{equation}
 p(z) \propto z^\alpha \exp\left[-\left(z/z_0\right)^\beta\right]\,,
\end{equation}
where we fix the parameters $\alpha$, $\beta$, and $z_0$ so that the median redshift of the 
model distribution reproduces the median redshift of the Kilo-Degree Survey 
\citep[KiDS,][]{dejong13}, $z_{\rm med}\simeq0.7$ (Kuijken et al., in prep). We split the lens 
sample in redshift bins of half-width $\Delta z=0.1$ to illustrate the results obtained from a 
tomographic cosmic shear analysis \citep[e.g.,][]{heymans13}. We use a narrow redshift bin covering 
$0.6<z<0.8$ for the GG and II power spectra, since this range is close to the one that maximizes 
the lensing signal in a KiDS-like tomographic analysis. The GI power spectrum is better captured by 
cross-correlating this redshift bin with one at low redshift; we choose $0.2<z<0.4$ as a compromise 
between a high intrinsic alignment efficiency and a large enough volume observed. 

\begin{figure}
 \centerline{\includegraphics[width=3.4in]{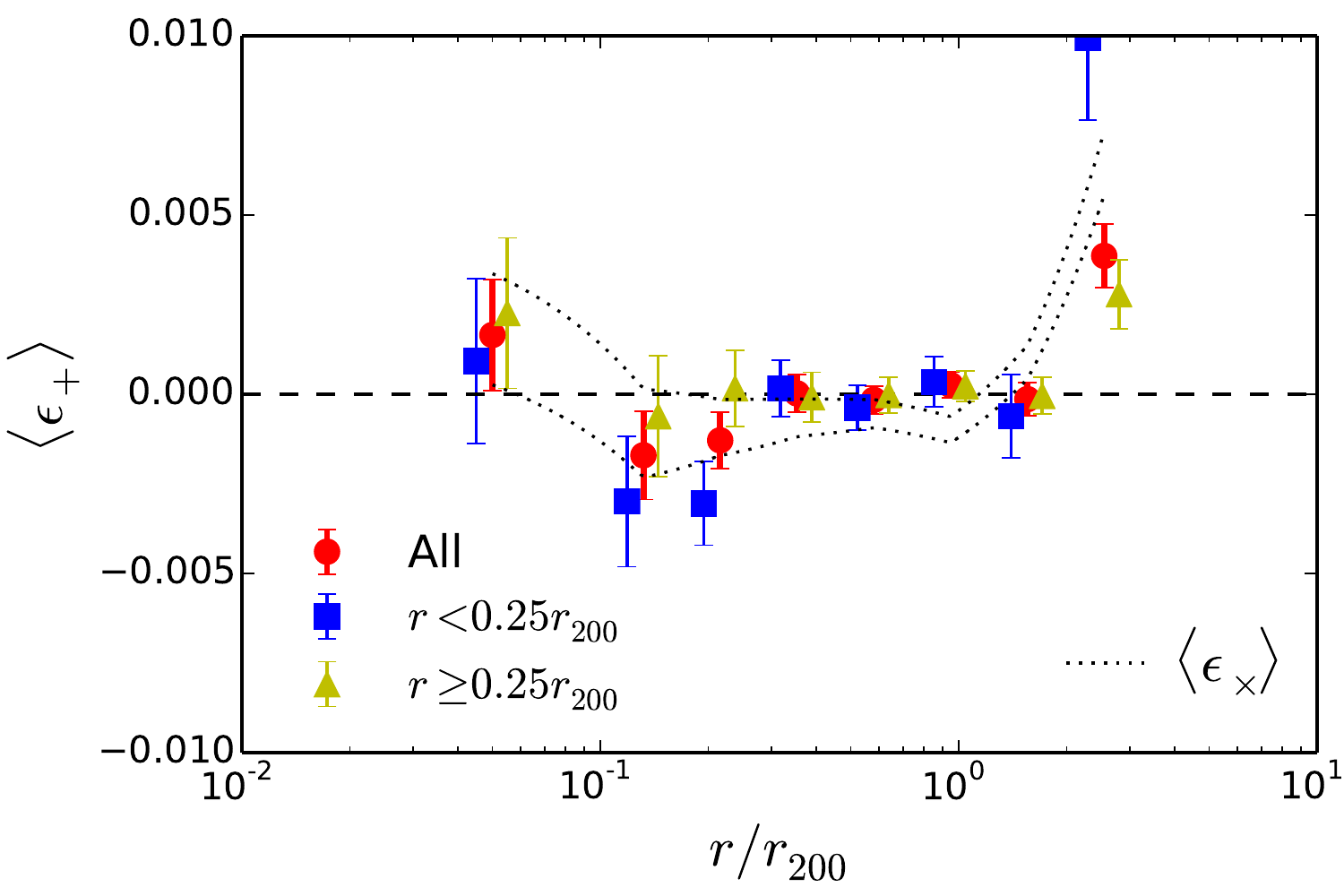}}
\caption{\small Satellite-satellite alignment as a function of the distance between satellites, 
for the full spectroscopic plus red sequence member sample. Red circles show all galaxies, while 
blue squares and yellow triangles show the signal with respect to galaxies inside and outside 
$0.25r_{200}$. Data points show the radial (positive) and tangential (negative) signal, while the 
dashed lines show the 68\% range of the cross component, linearly interpolated. Uncertainties do 
not account for covariance between data points. Note that the vertical scale is smaller than in
\cref{f:shear-all,f:shear-clsub,f:shear-galsub,f:bcg}.}
\label{f:satsat}
\end{figure}

\subsection{Linear alignment model}\label{s:linear}

The simplest models for galaxy alignments predict that elliptical galaxies are aligned with a 
strength that is proportional to the tidal field \citep{catelan01} while spiral galaxies, which are 
aligned by angular momentum acquired during gravitational collapse, are aligned with a strength 
that is proportional to the square of the tidal field \citep{pen00}. On sufficiently large scales, 
all galaxies are predicted to experience an alignment proportional to the large scale gravitational 
potential \citep{hui02}. Thus a linear alignment model is usually employed to characterize large 
scale galaxy alignments \citep[e.g.,][]{kirk10,joachimi11,mandelbaum11,heymans13}.\footnote{This 
model is typically referred to as ``nonlinear alignment model.'' However, this is a misnomer, 
since intrinsic alignments are still modeled as depending linearly on the tidal field; instead the 
name arises from the use of the nonlinear power spectrum in \cref{eq:EBspectraDef}. We therefore 
refer to it as linear alignment model throughout.} We normalize the intrinsic alignment power 
spectra as in previous studies \citep{hirata04,bridle07,schneider10}, matching to the SuperCOSMOS 
measurements of \cite{brown02}. This normalization is also consistent with more recent observations 
\citep{heymans04,mandelbaum06,joachimi11}.

Solid lines in \cref{f:powerspectra} show the angular power spectra, $C_\ell$, from the linear 
alignment model. This model includes no contribution from alignments within halos (so-called 1-halo 
terms) and therefore the II and GI power spectra are subdominant to the matter power spectrum at 
all 
scales.

\cref{f:powerspectra} also shows the expected angular power spectrum measurements of a reference 
cosmic shear survey with properties similar to KiDS with a redshift distribution as described 
above, with a sky coverage of 1,500 sq.\ deg.\ and a background source density of $n_{\rm 
gal}=10\,\mathrm{arcmin^{-2}}$. We assume a coverage $30\leq\ell\leq3000$, and compute the expected 
$C_\ell$ measurements and uncertainties following \cite{cooray01}, in logarithmic bins in $\ell$. 
The bottom panel of \cref{f:powerspectra} shows that the II contribution remains safely subdominant 
to statistical uncertainties expected for KiDS, but the GI contribution cannot be ignored, 
contaminating the GG power spectrum at the $\sim10\%$ level.

\subsection{Halo model}

The linear alignment model aims to describe alignments at large scales and the alignments 
between central galaxies, because these are expected to be aligned with the host halo by the large 
scale gravitational potential. On smaller, nonlinear scales, galaxy formation will tend to misalign 
baryonic and dark matter \citep[e.g.,][]{pereira10,tenneti14}, so the large-scale results from 
$N$-body simulations are probably not directly applicable to galaxy alignments within halos. 
Galaxy formation can also have a major impact on the power spectra \citep{vandaalen11,semboloni11}, 
and the way these two effects interplay is unclear. We therefore require a prescription to predict 
the power spectra accounting for 1-halo term galaxy alignments. To this end, we employ the halo 
model of radial alignments introduced by \cite{schneider10}.

The main assumption of the halo model is that galaxies form and reside in dark matter haloes whose 
masses directly influence the (observable) properties of the galaxies they host. Additionally, 
one can assume that satellite galaxies in a halo are radially aligned toward the center with a 
strength that can in principle be a function of the galaxy position in the halo, the host halo 
mass, and redshift. This is known as a \textit{satellite radial alignment model}. The total 
alignment can be separated into a prescription for galaxies in halos (the 1-halo term), and 
one between halos (the 2-halo term). We assume that galaxies populate halos following the halo 
occupation distribution of \cite{cacciato13} and the halo mass and bias functions of 
\cite{tinker10}. More details about the ingredients of this halo model can be found in 
\cite{schneider10}. Given a model for radial alignments, $\gamma^I(\mathbf{r},M,z)$, we calculate 
the power spectra through \cref{eq:EBspectraDef}.

This model only incorporates the information about the radial alignments studied in 
\cref{s:radial}, by definition. In principle, it would be possible to include further constraints 
on 
the alignments from measurements such as those explored in \cref{s:bcgalignment,s:satsat}. However, 
these would be second order corrections to the cluster-scale radial component. In particular, the 
satellite-satellite alignment constraints woud be relevant on scales smaller than what will be 
probed by current and upcoming experiments; we therefore choose not to include them in the present 
analysis.

\subsection{Impact of alignments within halos on the power spectra}\label{s:1haloIA}

\begin{figure}
 \centerline{\includegraphics[width=3.5in]{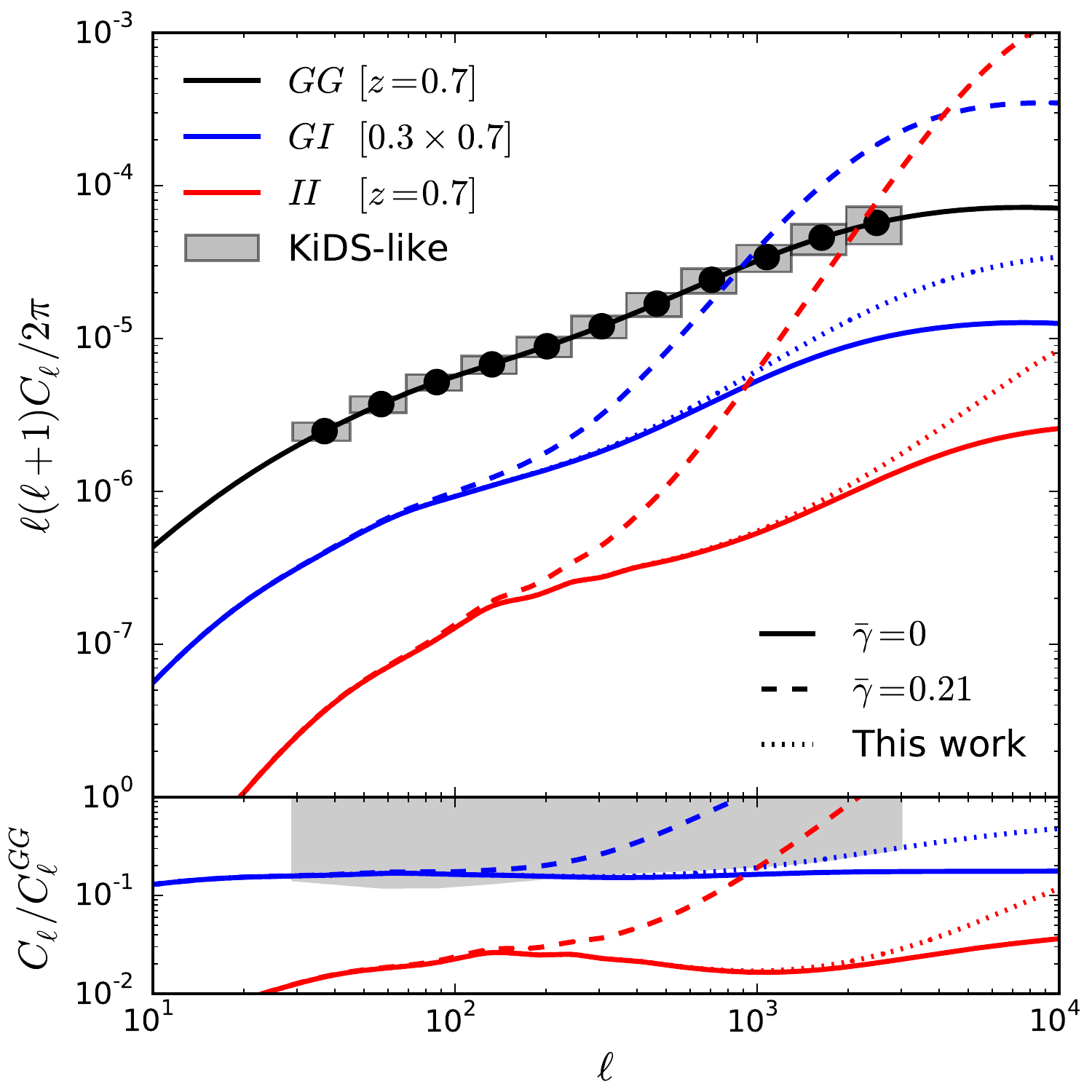}}
\caption{\small Effect of intrinsic alignments on the angular power spectra. \textit{Top panel:} 
The black line shows the GG power spectrum, while blue and red lines show the GI and II power 
spectra, respectively. Solid lines show the linear alignment model with no small-scale intrinsic 
alignments, while dashed and dotted lines model the contribution of satellite galaxies with 
$\ebar=0.21$ as in \cite{schneider10} and with the mass-dependent $2\sigma$ upper limit on the 
alignment signal derived in this work (see \cref{s:1haloIA}), respectively. Grey boxes with black 
circles show the expected uncertainty levels on a KiDS-like survey covering 1,500 sq.\ deg.\ and 
with $n_{\rm gal}=10\,\mathrm{arcmin^{-2}}$. The \textit{bottom panel} shows the ratio between the 
GI and II power spectra and the GG power spectrum, for each model. The shaded region shows values 
above the $1\sigma$ uncertainties in the anglar power spectrum for a KiDS-like survey, where GI and 
II contributions would dominate over statistical uncertainties.}
\label{f:powerspectra}
\end{figure}

The halo model requires a prescription for the strength of small-scale radial alignments. In its 
simplest form this strength is constant with radius and halo mass. The power spectra derived from 
this model are shown as dashed lines in \cref{f:powerspectra}, for an alignment strength 
$\ebar=0.21$ \citep[$\ebar$ is the 3-dimensional alignment strength derived from a projected 
measurement, $\gamma^I$; see][]{schneider10}. This is the fiducial value adopted by 
\cite{schneider10}. In this work, we extend this prescription by assuming a radial alignment that 
depends on halo mass but not on distance within the halo. Such a model is fully consistent with our 
results, since we find a null signal at all radii.

We construct a mass-dependent alignment model using the present results, plus the intrinsic 
alignment measurements in galaxy groups from the Galaxy and Mass Assembly (GAMA) survey 
\citep{schneider13}. We assume a power law for $\gamma^I(M)$, such that the mean ellipticity of 
satellite galaxies has the $2\sigma$ upper limits obtained in this study. We use the results for 
the augmented spectroscopic plus red sequence member sample, and choose to use the KSB measurement 
because, although the GALFIT constraint is less tight (i.e., more conservative), the contribution 
of each mass scale is weighted by the mass function. Since the mass function is an exponential 
function of mass, the overall alignment signal is dominated by lower mass objects. Since we use the 
constraint found for GAMA groups as a pivot, a smaller alignment strength in clusters will mean a 
larger overall contribution of alignments to a cosmic shear survey. Specifically, we use 
$\epsilon_+<0.0019$ at $M\simeq10^{15}\,M_\odot$ and $\epsilon_+<0.019$ at a typical mass 
$M\simeq10^{13}\,M_\odot$, corresponding to the $2\sigma$ upper limit for GAMA groups with $N_{\rm 
gal}\geq5$ \citep{schneider13}. Our model is therefore $\gamma^I(M)=(M/M_0)^{\alpha}$, constant 
with redshift, with $M_0 = 1.19\times10^9M_\odot$ and $\alpha=-0.5$.\footnote{The conversion 
between ellipticity and shear is given by $\gamma=\epsilon_+/2\mathcal{R}$, where $\mathcal{R}$ is 
the shear responsivity, which we assume to be equal to 0.87.} We note that the assumption of a 
single power law at all masses has no justification other than its simplicity. A more detailed halo 
model for intrinsic alignments will be presented in a forthcoming study (Cacciato et al., in prep), 
where we explore the impact of halo model assumptions on the predictions of the II and GI power 
spectra.

Dotted lines in \cref{f:powerspectra} show the intrinsic alignment power spectra predicted by the 
halo model for our adopted $\gamma^I(M)$. Since we constructed the model using $2\sigma$ upper 
limits on the measured alignments, the regions between the solid and dotted lines should be regarded 
as conservative estimates of the current uncertainties on the GI and II power spectra due to 1-halo 
term intrinsic alignments. As can be seen, both the GI and II power spectra remain subdominant to 
the GG power spectrum, which is not the case with the fiducial $\ebar=0.21$ model used by 
\cite{schneider10}. The GI angular power spectrum including our 1-halo term is $\sim$70\% higher 
than that predicted by the linear alignment model at $\ell\sim3000$, which translates into an excess 
on the total (GG+GI+II) angular power spectrum of $\approx10\%$, comparable to the statistical 
uncertainties expected at these scales. Note that at larger scales the GI power spectrum is 
dominated by linear alignments and the satellite contribution is well below the statistical 
uncertainties of KiDS. Therefore, we do not expect that cosmic shear analyses with KiDS will need to 
include a contribution by satellite galaxies in the modeling of intrinsic alignments, and we 
conclude that the linear alignment model should be a sufficient treatment of intrinsic 
alignments for KiDS. We note that for a bin at $0.2<z<0.4$, the II power spectrum can be $>10\%$ of 
the lensing (GG) power spectrum at the same redshift, but the uncertainties of a KiDS-like survey 
are much larger than at $z\sim0.8$ because of the smaller volume probed. In any case, the linear 
alignment model captures any II contribution to sufficient accuracy. Therefore a treatment of 
intrinsic alignments in KiDS cosmic shear analyses can rely on the linear alignment model, similar 
to the cosmic shear analysis of CFHTLenS data by \cite{heymans13}. We expect the situation to be 
similar for the Dark Energy Survey (DES)\footnote{\url{http://www.darkenergysurvey.org}}, which 
will have three times as much area as KiDS but otherwise similar characteristics. This may not be 
the case for larger surveys, for which the contribution of satellite galaxies to intrinsic 
alignments must be characterized to higher precision.

\section{Conclusions}

We have compiled a large sample of galaxies with spectroscopic redshifts in the direction of \Ncl\ 
galaxy clusters in the redshift range $0.05<z<0.55$, selected as part of MENeaCS and CCCP. We 
select cluster members using the shifting gapper technique, which uses phase space information, for 
a total \Nmembers\ cluster members. We use these members to estimate dynamical masses using the 
simulation-based scaling relation between velocity dispersion and cluster mass of \cite{evrard08}. 
The sample has a median redshift $z=0.14$ and a median mass $M_{200}\sim7\times10^{14}\,M_\odot$, 
in good agreement with the weak lensing masses estimated by \cite{hoekstra12}.

We quantify the alignment of galaxies within clusters using 14,250 cluster members for which 
we are able to measure their shapes either with KSB or GALFIT, after showing that the ellipticities 
measured by both methods are consistent (\cref{f:methods}). Both methods take different approaches 
to measuring galaxy shapes and therefore provide an important consistency check. We confirm that our 
analysis is free of significant systematic effects by measuring the average alignment of both 
foreground galaxies and stars. The signal from foreground galaxies is consistent with zero; the 
signal from stars is significantly different from zero, but at a level of 
$\langle\epsilon_+\rangle\sim10^{-4}$, an order of magnitude lower than measurement uncertainties 
(\cref{f:control}).

We measure three different alignments: the radial alignment of satellite galaxies toward the BCG, 
the alignment of satellites with the BCG orientation, and the radial alignment of satellites toward 
each other. Each probes a different, but not necessarily independent, effect. We find no evidence 
for any of these alignments (\cref{f:shear-all,f:shear-clsub,f:shear-galsub,f:bcg,f:satsat}). In 
particular, we constrain the average ellipticity of satellites toward BCGs to 
$\langle\epsilon_+\rangle=-0.0037\pm0.0027$ with KSB and 
$\langle\epsilon_+\rangle=0.0004\pm0.0031$ with GALFIT, at 68\% confidence, within \radius. 
Similarly, there is no evidence of galaxy alignments when splitting the sample by cluster (redshift, 
mass, or dynamical state) or galaxy (color or luminosity) properties. Selecting additional cluster 
members through the red sequence allows us to extend the sample to $\sim$20,000 galaxies with an 
estimated contamination of $<10\%$ from red sequence interlopers (\cref{f:rspurity}). All signals 
from this enlarged sample are also consistent with zero.

We include this constraint on the radial alignment of galaxies within high-mass halos, together 
with a measurement at the group scale \citep{schneider13}, in a halo model framework, and derive 
the current uncertainty on the angular power spectrum given by intrinsic alignments within halos (a 
1-halo term). We find that the total (GG+GI+II) angular power spectrum predicted from our alignment 
model (see \cref{s:1haloIA}) is, at most, 10\% higher than the total power spectrum predicted by the 
linear alignment model at the smallest scales probed by KiDS, $\ell\sim3000$. This level of 
contribution from satellite galaxies will not be relevant for cosmic shear measurements with KiDS or 
DES (see \cref{f:powerspectra}). We conclude that the linear alignment model is a sufficient 
description of intrinsic alignments for KiDS, but the situation may be different for significantly 
larger surveys.

\begin{acknowledgements}

We thank Chris Pritchet and Dennis Zaritsky for help carrying out the MENeaCS spectroscopic survey, 
Patrick Kelly for making his python SLR code publicly available, and Radek Wojtak for suggesting an 
assessment of orientation bias.
CS, HH, and MV acknowledge support from the European Research Council under FP7 grant number 
279396.
HH, MC and RFJvdB acknowledge support from the Netherlands Organisation for Scientific Research 
grant number 639.042.814.

Based on observations obtained with MegaPrime/MegaCam, a joint project of CFHT and CEA/IRFU, at the 
Canada-France-Hawaii Telescope (CFHT) which is operated by the National Research Council (NRC) of 
Canada, the Institut National des Science de l'Univers of the Centre National de la Recherche 
Scientifique (CNRS) of France, and the University of Hawaii. 

Funding for SDSS-III has been provided by the Alfred P. Sloan Foundation, the Participating
Institutions, the National Science Foundation, and the U.S. Department of Energy Office of Science.
The SDSS-III web site is http://www.sdss3.org/.

This research has made use of the NASA/IPAC Extragalactic Database (NED) which is operated by the 
Jet Propulsion Laboratory, California Institute of Technology, under contract with the National 
Aeronautics and Space Administration.

This publication makes use of data products from the Two Micron All Sky Survey, which is a joint 
project of the University of Massachusetts and the Infrared Processing and Analysis 
Center/California Institute of Technology, funded by the National Aeronautics and Space 
Administration and the National Science Foundation.

This research has made use of NASA's Astrophysics Data System Bibliographic Services.

This work has made use of the python packages \texttt{scipy} (http://www.scipy.org) and 
\texttt{matplotlib} \citep[http://matplotlib.org;][]{hunter07} and of IPython \citep{perez07}.

\end{acknowledgements}

\end{document}